\documentclass[11pt,a4]{article}

\pdfoutput=1

\usepackage{jheppub}

\usepackage{graphicx,amssymb,amsmath,amsfonts}

\title{Glueballs as rotating folded closed strings}

\preprint{TAUP-2995-15}

\author{Jacob Sonnenschein and Dorin Weissman}

\affiliation{The Raymond and Beverly Sackler School of Physics and Astronomy,\\
	Tel Aviv University, Ramat Aviv 69978, Israel}

\emailAdd{cobi@post.tau.ac.il}
\emailAdd{dorinw@mail.tau.ac.il}

\abstract{
In previous papers \cite{Sonnenschein:2014jwa,Sonnenschein:2014bia} we argued that mesons and baryons can be described as rotating open strings in holographic backgrounds. Now we turn to closed strings, which should be the duals of glueballs. We look at the rotating folded closed string in both flat and curved backgrounds.

A basic prediction of the closed string model is that the slope of Regge trajectories is half that of open strings. We propose that a simple method to identify glueballs is to look for resonances that belong to trajectories with a slope of approximately 0.45 GeV$^{-2}$, half the meson slope. We therefore look at the experimental spectra of flavorless light mesons to see if such a scheme, where some of the states are placed on open string trajectories and some on closed ones, can fit known experimental data. We look at the $f_0$ ($J^{PC} = 0^{++}$) and $f_2$ ($2^{++}$) resonances. As there is no preference for a single scheme of sorting the different states into meson and glueball trajectories, we present several possibilities, each identifying a different state as the glueball. We supplement each scheme with predictions for the masses of excited glueballs.

We show that the width of the decay into two mesons is different for glueballs and mesons thus providing a supplementary tool to distinguish between them. In addition, we look at some lattice QCD results for glueball spectra and check their compatibility with the closed string model.

One of the main conclusions of this paper is that an extension of experimental data on the spectrum of flavorless hadrons is needed, in particular in the region between around 2.4 GeV and 3 GeV.
}


\def\be{\begin{equation}}
\def\ee{\end{equation}}

\newcommand{\alp}{\ensuremath{\alpha^\prime}}

\newcommand{\ssb}{\ensuremath{s\bar{s}}}

\newcommand{\plm}{\ensuremath{\pm}}

\newcommand{\bea}{\begin{eqnarray}}
\newcommand{\eea}{\end{eqnarray}}

\begin{document}
\maketitle
\flushbottom

\section{Introduction}

A well known item of the string/gauge theory holographic dictionary states that closed strings are the duals of glueballs in the corresponding gauge theories. On the other hand, using the gravity/gauge theory duality, glueball operators of the boundary field theory correspond to fields in the gravity bulk theory, in particular modes of the dilaton, the graviton, and the RR field. Using this latter correspondence a spectrum of holographic glueballs has been determined \cite{Csaki:1998qr,Brower:2000rp,Elander:2013jqa}. However, from the same reasoning as in \cite{Sonnenschein:2014jwa} and \cite{Sonnenschein:2014bia}, it is probably the spectrum of the strings and not of the bulk fields that would really correspond to the experimental data when moving from large $N_c$ and large $\lambda$ to the to the realistic values of $N_c=3$ and $\lambda\sim 1$. For  mesons \cite{Sonnenschein:2014jwa} and baryons \cite{Sonnenschein:2014bia} it was argued that the (open) string configurations admit a modified Regge behavior that matches that of the observed hadrons whereas bulk modes do not admit this property. In this paper we argue that a similar correspondence exists for glueballs, and that the glueballs will probably have a better description in terms of closed strings rather than modes of bulk fields. The idea of glueballs as closed strings has previously been discussed in various terms in works such as \cite{Bhanot:1980fx,Niemi:2003hb,Sharov:2007ag,Solovev:2000nb,Talalov:2001cp,LlanesEstrada:2000jw,Szczepaniak:2003mr,Pons:2004dk,Abreu:2005uw, Brau:2004xw,Mathieu:2005wc,Simonov:2006re,Mathieu:2006bp,BoschiFilho:2002vd}.

It is common lore that glueballs and flavorless mesons cannot be distinguished since they carry the same quantum numbers and that the corresponding resonances encountered in experiments are in fact generically linear combinations of the two kinds of states. If, however, we refer to the stringy description of hadrons then, since mesons and glueballs correspond to open and closed strings respectively, there are certain characterizing features with which one can distinguish between them.

The most important difference between the open string mesons and the closed string glueballs is the slope, or equivalently the (effective) tension. It is a basic property of strings (see sections \ref{sec:lightcone} and \ref{sec:flat_string}) that the effective tension of a closed string is twice that of an open string and hence there is a major difference between the two types of strings, as
\be \alp_{closed}= \frac12 \alp_{open}\qquad  \rightarrow \qquad \alp_{gb}= \frac12 \alp_{meson} \,. \ee
Thus the basic idea of this paper is that one should be able to distinguish between glueballs and flavorless mesons by assigning some of them to certain trajectories with a mesonic slope $\alp_{meson}\sim 0.9$ GeV$^{-2}$ and others to trajectories with a glueball slope of $\alp_{gb}\sim 0.45$ GeV$^{-2}$.

The slope is not the only thing that is different between open and closed strings. It follows trivially from the spectrum of closed strings (see section \ref{sec:lightcone}) that in the critical dimension it has an intercept that is twice that of the open string. However, we are interested in strings in four dimensions rather than the critical dimension, and there, as will be discussed in section \ref{sec:non_critical_string}, the determination of the intercept is still not fully understood. Thus the intercept cannot currently serve as a tool for identifying glueballs.

Another important difference between open and closed string hadrons is in their decay mechanisms. Based on the holographic description of a meson as a string that connects to flavor branes at its two endpoints, it was determined in \cite{Peeters:2005fq} that the width of decay of a meson of mass \(M\) behaves like
\be \Gamma \sim \left(\frac{ 2 M}{\pi T} - \frac{m^1_{sep}+m^2_{sep}}{2T}\right)  e^{\frac{-{m^q_{sep}}^2}{T}} \,, \label{eq:holo_width}\ee
where $M$ is the mass of the meson, $m^1_{sep}$ and $m^2_{sep}$ are the masses of the string endpoint quarks in the initial state (assumed here to be small), $m^q_{sep}$ is the mass of the quark and antiquark pair generated by the split of the string and $T$ is the string tension. The factor preceding the exponent is in fact the string length.

As we discuss in section \ref{sec:decays}, for the case of a closed string decaying into two open strings the width will be proportional to the string length squared, and the single exponential suppressing factor will be replaced by
\be e^{\frac{-m^q_{sep}{}^2}{T}}e^{\frac{-m^{q^\prime}_{sep}{}^2}{T}} \,,\ee
where $m^{q}_{sep}$ and $ m^{q^\prime}_{sep}$ are the masses of each of the two quark-antiquark pairs that will have to be created in the process.

Thus it is clear that the width of a glueball should be narrower than that of the corresponding meson open string, particularly for decay channels involving heavier quarks like $s$, $c$ and $b$. This can serve as an additional tool of disentangling between mesons and glueballs. We list one distinguishing feature of a glueball decaying into two mesons in section \ref{sec:decays}.

The main motivation of reviving the description of mesons and baryons in terms of open strings in \cite{Sonnenschein:2014jwa} and \cite{Sonnenschein:2014bia} has been the holographic string/gauge duality. The same applies also to the closed string picture of glueballs.   The spectra of closed strings in a class of holographic confining models was analyzed in \cite{PandoZayas:2003yb}. The result was that the relation between the mass and angular momentum takes the following form:
\be J = \alp_{gb} (E^2 - 2 m_0 E) + a \,, \ee
where $\alp_{gb}$ is the corresponding slope, $E$ is the mass of the glueball, $a$ is the intercept, and $m_0$ is a parameter that can be either positive or negative and is determined by the particular holographic model used. Note that this relation modifies the well known linear relation between $J$ and $E^2$. In section \ref{sec:holo_fits} we discuss the phenomenological implications of this relation and analyze the possibility of grouping flavorless hadrons along such holographic trajectories.

The main goal of this paper is to perform an explicit comparison between observational data of flavorless hadrons and the resonance states predicted by the models of rotating open string with massive endpoints for the mesons and rotating folded closed strings for glueballs.

Unfortunately there exists no unambiguous way to assign the known flavorless hadrons (the focus in this paper is on the \(f_0\) and \(f_2\) resonances) into trajectories of mesons and glueballs, but it is clear that \textbf{one cannot consistently sort all the known resonances into meson trajectories alone}. One of the main problems in identifying glueball trajectories is simply the lack of experimental data, particularly in the mass region between \(2.4\) GeV and the \(c\bar{c}\) threshold, the region where we expect the first excited states of the glueball to be found. It is because of this that we cannot find a glueball trajectory in the angular momentum plane.

We mostly focused then on the radial trajectories of the \(f_0\) (\(J^{PC} = 0^{++}\)) and \(f_2\) (\(2^{++}\)) resonances. For the \(f_0\) we examined the possibility of identifying one of the states \(f_0(980)\), \(f_0(1370)\), \(f_0(1500)\), or \(f_0(1710)\) as the glueball ground state and building the trajectories beginning from those states. This procedure did not show any significant preference for any one of the glueball candidates over the other. For the \(f_2\) there is less ambiguity, but still no positive identification. Between the different \(2^{++}\) state we find that the two very narrow resonances \(f_2(1430)\) and \(f_J(2220)\) (the latter being a popular candidate for the tensor glueball) do not belong on meson trajectories.

The paper is organized as follows. Section \ref{sec:theory} is devoted to the theory of rotating closed strings. In section \ref{sec:lightcone} we review the light-cone quantization of the basic bosonic string and describe its spectrum. Next we address the rotating folded string. We present the classical solution and the corresponding Regge trajectory, starting by discussing the case of flat spacetime. We introduce the Polchinski-Strominger term needed to assure two dimensional conformal invariance in non-critical dimension and discuss the problematic result for the intercept for a folded closed string in four dimensions. In section \ref{sec:holo_string} we review the results of \cite{PandoZayas:2003yb} for the rotating folded string in holographic backgrounds, and the semiclassical correction obtained there. Section \ref{sec:decays} is devoted to the decay process of string decaying into two strings. We summarize the result for the decay of an open string into two open strings \cite{Peeters:2005fq} and generalize it also to the case of a closed string decaying into two open strings. Section \ref{sec:phenomenology} deals with the phenomenology of the rotating folded string models and the comparison between them and the observational data. We begin by spelling out the basic assumptions of the phenomenological models in section \ref{sec:fitting_models}. We then present the key experimental players: the $f_0$ and $f_2$ resonances. In \ref{sec:f0_fits} we propose several assignments of the $f_0$ resonances into radial $(n,M^2)$ trajectories, first into only various mesonic trajectories and then into various possible combinations when singling out some states as glueballs. In \ref{sec:f2_fits} we describe possible assignments of the $f_2$, first into orbital $(J,M^2)$ trajectories, then into \((n,M^2)\) trajectories. Section \ref{sec:holo_fits} expands on previous sections by using the non-linear trajectory that characterizes the glueballs of holographic models. In section \ref{sec:lattice} we discuss the spectrum of glueballs that follows from lattice gauge theory models. We review the trajectories determined in lattice simulations and their corresponding slopes. Both types of trajectories, $(J,M^2)$ and $(n,M^2)$, are discussed. Section \ref{sec:summary} is a summary and discussion of the results and states some open questions. In the appendix \ref{sec:predictions} we list the predictions of our models for the yet unobserved excited partners of the glueball candidates, based on their Regge trajectories.

\section{The rotating closed string} \label{sec:theory}
\subsection{Quantized closed string in light cone gauge} \label{sec:lightcone}
We review here the derivation of the spectrum of the bosonic closed string in the light cone gauge. We simply present the derivation in chapter 1 of \cite{Polchinski:Vol1}, omitting some of the details for brevity's sake. The following treatment is essentially true only for the critical dimension \(D = 26\), but we keep a general \(D\) in the formulae. We return to this point in section \ref{sec:non_critical_string}.

We start from the Polyakov action
\be S = -\frac{1}{4\pi\alp}\int d\tau d\sigma \sqrt{-\gamma}\gamma^{\alpha\beta}\eta_{\mu\nu}
\partial_\alpha X^\mu \partial_\beta X^\nu \,.\ee
We define the light cone coordinates \(x^\pm = \frac{1}{\sqrt{2}}(x^0\pm ix^1)\), and set the gauge by making the three requirements
\be X^+ = \tau\,, \qquad \partial_\sigma \gamma_{\sigma\sigma} = 0\,, \qquad \sqrt{-\gamma} = 1  \label{eq:lightcone} \,.\ee
The equations of motion for the transverse coordinates are then simple wave equations and they are generally solved (with closed string boundary conditions, for \(\sigma \in (-\ell,\ell)\)) by
\be X^i(\sigma,\tau) = x^i + \frac{p^i}{p^+}\tau + i\left(\frac{\alp}{2}\right)^{1/2}
\sum_{n\neq0}\left[\frac{\alpha^i_n}{n}\exp\left(-i\frac{2\pi n (\sigma+c\tau)}{2\ell}\right)
 + \frac{\beta^i_n}{n}\exp\left(i\frac{2\pi n (\sigma-c\tau)}{2\ell}\right)\right] \,.\ee
The constant \(c\) is related to the coordinate length \(\ell\) and the conserved quantity \(p^+\) via \(c = \ell/(\pi\alp p^+)\). Aside from \(\ell\), which is proportional to the physical string length, these constants do not have any significance on their own except in keeping track of units.

The left and right moving modes, \(\alpha^i_n\) and \(\beta^i_n\), are independent of each other (and hence, commute) and are normalized in such a way that
\be [\alpha^i_m,\alpha^j_n] = [\beta^i_m,\beta^j_n] = m\delta^{ij}\delta_{m,-n} \,.\ee
The Hamiltonian has the mode expansion
\be H = \frac{p^i p^i}{2p^+} + \frac{1}{\pi\alp}\left[\sum_{n>0}\left(\alpha^i_{-n}\alpha^i_n+\beta^i_{-n}\beta^i_n\right)+ A + \tilde{A}\right] \,, \ee
noting that \((\alpha^i_n)^\dagger = \alpha^i_{-n}\). \(A\) and \(\tilde{A}\) are the c-numbers one gets when normal-ordering the sums. After regularizing the appropriate infinite sums, identical for the left and the right moving modes, and taking contributions from \(D-2\) transverse modes, we get the result
\be A = \tilde{A} = \frac{2-D}{24} \,.\ee

From here we get the spectrum using the mass shell condition \(M^2 = -p^2 = 2p^+ H - p^i p^i\), which translates to
\be M^2 = \frac{2}{\alp}\left(\sum_{n>0}\left(\alpha^i_{-n}\alpha^i_n+\beta^i_{-n}\beta^i_n\right)+ A + \tilde{A}\right) \,,\ee
or,
\be M^2 = \frac{2}{\alp}\left(N+\tilde{N}+A+\tilde{A}\right) \,,\ee
where \(N\) and \(\tilde{N}\) are the total population numbers of the left and right moving modes.

For comparison, the same treatment of the open string leads to the result
\be M^2_{open} = \frac{1}{\alp}\left(N+A\right) \,.\ee
Here we have neither the constant prefactor of two which halves the slope of the closed string, nor do we have two different kinds of modes on the string and the resulting doubling of the intercept.

\subsubsection{Quantized closed string: The spectrum}
While the left and right moving modes on the closed string are independent, there is one constraint that relates them, affecting the spectrum. After making the gauge choice by imposing the three conditions of eq. \ref{eq:lightcone} we still have a residual symmetry of \(\tau\)-independent translations of \(\sigma\). This results in the additional constraint
\be N = \tilde{N} \,.\ee
The total number of excitations has to be equal for the left and right moving modes.

The vacuum state of the closed string is defined as the state annihilated by all \(\alpha^i_n\) and \(\beta^i_n\), for positive \(n\). It has \(N = \tilde{N} = 0\), we denote it simply \(|0\rangle\),\footnote{The vacuum state may also have some center of mass momentum \(p\), but we suppress it in this notation.} and its mass is determined by the intercepts:
\be M^2 = \frac{2}{\alp}(A+\tilde{A}) = \frac{2-D}{6\alp} \,.\ee
For \(D = 26\) this state is a tachyon, with \(M^2 = -4/\alp\). The first excited state has \(N = \tilde{N} = 1\), and so is of the form
\be \alpha^{i}_{-1}\beta^j_{-1}|0\rangle \ee
and its mass is
\be M^2 = \frac{2}{\alp}(2+A+\tilde{A}) = \frac{26-D}{6\alp} \,.\ee
In the critical dimension we have here a massless tensor and a massless scalar.

The most important feature of the spectrum for our uses is that it forms an infinite tower of states, with the difference between each pair of consecutive states being
\be \Delta M^2 = \frac4\alp \,, \ee
with one factor of two coming from the halving of the slope, and the other from the fact that \(N+\tilde{N}\) takes only even values: \(0,2,4,6,\ldots\).

\subsection{The rotating closed string solution} \label{sec:rotating_string}
\subsubsection{Classical rotating folded string} \label{sec:flat_string}

Here we use the Nambu-Goto action for the string
\be S = -\frac{1}{2\pi\alp}\int d\tau d\sigma \sqrt{-h} \,,\ee
with
\be h = \det h_{\alpha\beta}\,, \qquad h_{\alpha\beta} = \eta_{\mu\nu}\partial_\alpha X^\mu \partial_\beta X^\nu \,,\ee
and \be \alp = \frac{1}{2\pi T} \,. \ee

The rotating folded string is the solution
\be X^0 = \tau \qquad X^1 = \frac{1}{\omega}\sin(\omega\sigma)\cos(\omega\tau) \qquad X^2 = \frac{1}{\omega}\sin(\omega\sigma)\sin(\omega\tau) \,.\label{eq:rotsol}\ee
We take \(\sigma \in (-\ell,\ell)\) and correspondingly \(\omega\) takes the value \(\omega = \pi/\ell\). The energy of this configuration is
\be E = T \int_{-\ell}^\ell d\sigma \partial_\tau X^0 = 2T\ell \,.\ee
The angular momentum we can get by going to polar coordinates (\(X^1 = \rho\cos\theta, X^2 = \rho\sin\theta\)), then
\be J = T \int_{-\ell}^\ell d\sigma \rho^2 \partial_\tau \theta =
\frac{T}{\omega} \int_{-\ell}^\ell d\sigma \sin^2(\omega\sigma) = \frac{\pi T}{\omega^2} = \frac{T\ell^2}{\pi}\,.\ee
From the last two equations we can easily see that for the classical rotating folded string
\be J = \frac{1}{4\pi T}E^2 = \frac{1}{2}\alp E^2 \,.\ee

\subsubsection{Quantization of the rotating folded string} \label{sec:non_critical_string}
In a previous section we reviewed the quantization of the bosonic closed string in the critical dimension, \(D = 26\). There we have the result
\be \frac{1}{2}\alp M^2 = N + \tilde{N} - \frac{D-2}{12} \,. \ee
We would like to obtain a correction to the classical trajectory of a similar form when quantizing the rotating folded string in \(D = 4\) dimensions. In \cite{Hellerman:2013kba} the intercept was computed in the context of effective string theory where the Polchinski-Strominger (PS) term \cite{Polchinski:1991ax},
\be \mathcal{L}_{PS} = \frac{26-D}{24\pi}\frac{(\partial_+^2X\cdot\partial_-X)(\partial_-^2X\cdot\partial_+X)}{(\partial_+X\cdot\partial_-X)^2} \label{eq:psterm} \,, \ee
compensates for the conformal anomaly when working outside the critical dimension. The derivatives are with respect to the variables \(\sigma^\pm \equiv \tau\pm\sigma\).

As was described in the introduction and will be further discussed in section \ref{sec:holo_string}, a major candidate for describing the glueball is a rotating closed string in a holographic background which lives, by definition, in the critical dimension. One may conclude that in this case the PS term is not needed. However, as was argued in \cite{Aharony:2009gg}, upon integrating out the massive degrees of freedom of the closed string that resides in the  critical holographic dimension one gets the PS action as part of the effective string action in the non-critical $D$ dimensions.  

The calculation in \cite{Hellerman:2013kba} is for a general dimension \(D\), with, as already mentioned, the PS term included. In dimensions larger than four the string will rotate in two planes and the angular momentum is characterized by two quantum numbers \(J_1\) and \(J_2\). The result obtained there for the Regge trajectory of the closed string is
\be \frac{\alp}{2}M^2 = (J_1+J_2) - \frac{D-2}{12} + \frac{26-D}{24}
\left((\frac{J_1}{J_2})^\frac{1}{4}-(\frac{J_2}{J_1})^\frac{1}{4}\right)^2 \,. \ee
This expression is singular when \(J_2 = 0\), which is necessarily the case when \(D = 4\), since in four dimensions the rotation is in a single plane. Therefore the expression is not usable precisely in the context in which we would like to use it.

We can see where this originates by inserting the 4D rotating solution from eq. \ref{eq:rotsol} into the expression for the PS term, eq. \ref{eq:psterm}. The expression obtained,
\be \mathcal{L}_{PS} = -\frac{D-26}{24\pi}\omega^2\tan^2(\omega\sigma) \,,\ee
is singular when \(\omega\sigma = \pm\frac{\pi}{2}\), i.e. at the two points \(\sigma = \pm\frac{\ell}{2}\), which are the ``endpoints'', or folding-points, of the rotating folded string, and the integral on \(\mathcal{L}_{PS}\) giving the correction diverges:
\be \int_{-\ell}^\ell d\sigma \mathcal{L}_{PS} = -\frac{D-26}{12\pi}\omega^2\int_{-\ell/2}^{\ell/2}d\sigma\tan^2(\omega\sigma) = -\frac{D-26}{12\pi}\omega\left(\tan x-x\right)|_{x=-\pi/2}^{\pi/2} \,.\ee
We see that beneath the divergent \(\tan x\) there is also the finite part
\be \frac{D-26}{12}\frac{\pi}{\ell} \,.\ee

The denominator in the PS term is simply \((\dot{X}^2)^2\), so the problem emerges because the endpoints move at the speed of light. The same problem is encountered in the treatment of the open string, but as was shown in \cite{Hellerman:2013kba} in that case one can introduce a counterterm at the string boundaries that renders the action and correspondingly the intercept finite. In fact it was found out that summing up the contributions to the latter from the PS and from the Casimir term, the \(D\) dependence is canceled out between the two terms, and the intercept is given simply by $a= 1$, for all \(D\). Another possible approach for regularizing the rotating open string is to add masses to its endpoints. However, the quantization of the system of a rotating string with massive particles on its ends is still not fully understood \cite{ASY}.

For the closed string it is not clear how to regularize the system. One potential way to do it might be to add two masses at the two endpoints of the folded string. The resulting system looks like two open strings connected at their boundaries by these masses, but not interacting in any other way. In the rotating solution the two strings lie on top of one another. The boundary condition, which is the equation of motion of the massive endpoint is modified: it is the same as for the open string, but with an effective double tension \(T \rightarrow 2T\), in accordance with the ratio of the slopes of the open and closed strings discussed above. In fact everything else is doubled too. If this process of adding masses on the closed string and taking then the limit of zero mass is a legitimate way to regularize, then it is probable that the result is also simply double that of the open string, as it is for the critical dimension. Obviously, though, even in that case we cannot perform the quantization of the folded closed string since, as mentioned above, we do not fully control the quantization of an open string with massive endpoints.   

\subsubsection{The closed string in a curved background} \label{sec:holo_string}
The full analysis of rotating closed string in holographic curved backgrounds was performed in \cite{PandoZayas:2003yb}. We present here the key points in short form.

\begin{figure}[t!] \centering
	\includegraphics[width=0.95\textwidth]{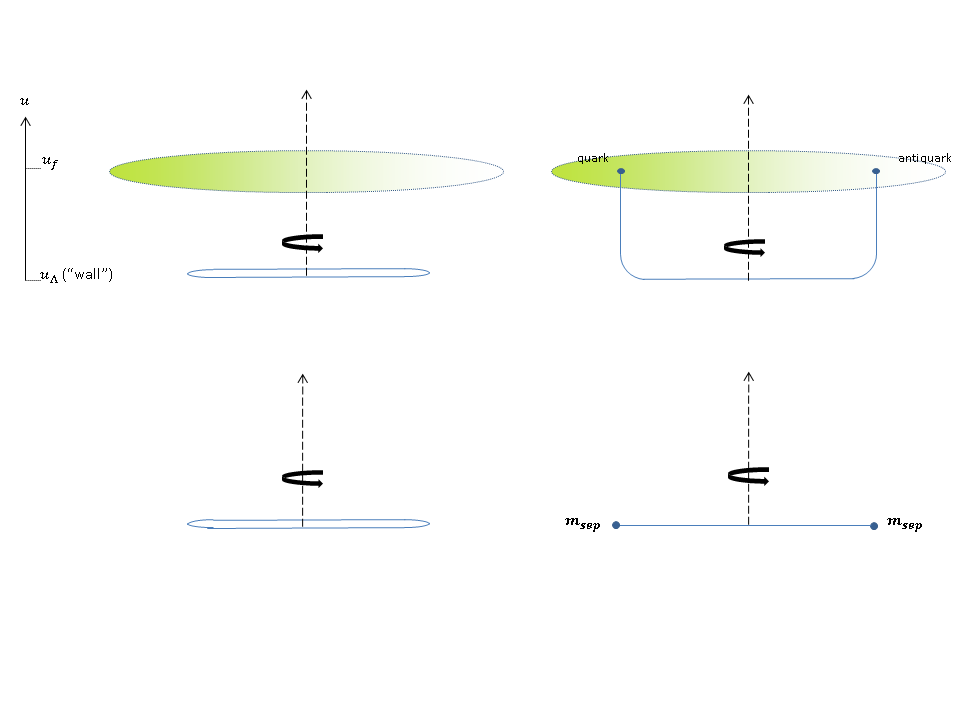} \\
	\caption{\label{fig:closed_open_map} Closed and open strings in holographic backgrounds (top), and their mappings into flat spacetime (bottom). For the open string, the mapping from curved to flat background adds endpoint masses to the strings \cite{Kruczenski:2004me,Sonnenschein:2014jwa}, with the vertical segments mapped to the point-like masses in flat space. For the closed string, we look at the simple folded string in both cases. Note that classically the rotating folded string has zero width, and as such would look like an open string with no endpoint masses, and not like in the drawing.}
			\end{figure}

If we look at a curved background metric of the form
\be ds^2 = h(r)^{-1/2}(-dX^0dX^0+dX^idX^i) + h(r)^{1/2}dr^2 + \ldots \,,\ee
with \(i = 1,2,3\) and the ellipsis denoting additional transverse coordinates, the rotating folded string, namely the configuration,
\be X^0 = l\tau \qquad X^1 = l\sin\sigma\cos\tau \qquad X^2 = l\sin\sigma\sin\tau \,,\ee
is still\footnote{We follow a somewhat different normalization here, taking \(\omega = \pi/\ell\) from the previous section to be \(1\), and introducing a common prefactor \(l\), but the solution is essentially the same as the flat space solution of section \ref{sec:flat_string}.} a solution to the string equations of motion provided we take
\be r(\sigma,\tau) = r_0 = Const. \ee
where \(r_0\) is a point where the metric satisfies the condition
\be \partial_r g_{00}(r)|_{r=r_0} = 0, \qquad g_{00}(r)|_{r=r_0} \neq 0 \,.\ee
The existence of such a point is also one of the sufficient conditions for the dual gauge theory to be confining \cite{Kinar:1998vq}. Compared to the folded string in flat spacetime, the energy and angular momentum take each an additional factor in the form of \(g_{00}(r_0)\):
\be E = \frac{1}{2\pi\alp}\int_{-\pi}^\pi g_{00}(r_0)d\sigma = g_{00}(r_0) \frac{l}{\alp} \,,\ee
\be J = T\int_{-\pi}^\pi g_{00}(r_0)\sin^2\sigma d\sigma = g_{00}(r_0) \frac{l^2}{2\alp} \,.\ee
Defining an effective string tension \(T_{eff} = g_{00}(r_0)T\) and slope \(\alp_{eff} = (2\pi T_{eff})^{-1}\), we can write the relation
\be J = \frac{1}{2}\alp_{eff} E^2 \,.\ee
The same factor of \(g_{00}(r_0)\) multiplies the effective tension in the open string case, and therefore the closed and open string slopes are still related by the factor of one half, although the open string trajectories have the additional modification which can be ascribed to the presence of endpoint masses \cite{Kruczenski:2004me,Sonnenschein:2014jwa}. We draw the two types of strings in figure \ref{fig:closed_open_map}.

Calculations of the quantum corrected trajectory of the folded closed string in a curved background in different holographic backgrounds were performed in \cite{PandoZayas:2003yb} and \cite{Bigazzi:2004ze} using semiclassical methods. This was done by computing the spectrum of quadratic fluctuations, bosonic and fermionic, around the classical configuration of the folded string.
It was shown in \cite{PandoZayas:2003yb} that the Noether charges of the energy  $E$ and angular momentum  $J$ that incorporate the quantum fluctuations, are related to the expectation value of the world-sheet Hamiltonian in the following manner:
\be
lE -J = \int d\sigma <{\cal H}_{ws}> \,.
\ee
The contributions to the expectation value of the world-sheet  Hamiltonian are from several massless bosonic modes, ``massive" bosonic modes and massive fermionic modes.  
 For the ``massive" bosonic fluctuations around the rotating solution one gets a \(\sigma\)-dependent mass term, with equations of motion of the form
\be (\partial_\tau^2-\partial_\sigma^2+2m_0^2l^2\cos^2\sigma)\delta X^i = 0 \ee
appearing in both analyses, \(m_0\) being model dependent. A similar mass term, also with \(\cos\sigma\), appears in the equations of motion for some fermionic fluctuations as well, the factor of \(\cos^2\sigma\) in the mass squared coming in both cases from the induced metric calculated for the rotating string, which is \(h_{\alpha\beta} \sim \eta_{\alpha\beta}\cos^2\sigma\).

The result in both papers is that the Regge trajectories are of the form
\be J = \alp_{closed}(E^2- 2m_0 E) +a  \,.\ee
where $m_0$ is a mass parameter that characterizes the holographic model and $a$ is the intercept which generically takes the form $a= \frac{\pi}{24}(\#\text{bosonic massless  modes} - \#\text{fermionic massless modes})$. The two papers \cite{PandoZayas:2003yb} and \cite{Bigazzi:2004ze} use different holographic models (Klebanov-Strassler and Maldacena-N\'{u}\~{n}ez backgrounds in the former and Witten background in the latter) and predict different signs for \(m_0\), which is given as a combination of the parameters specific to the background. In \cite{PandoZayas:2003yb} \(m_0\) is positive, while in \cite{Bigazzi:2004ze} it is negative. According to \cite{PandoZayas:2003yb} the slope of the closed string trajectory is left unchanged from the classical case
\be \alp\!_{closed} = \frac{1}{2}\alp\!_{open} \,,\ee
while the model used in \cite{Bigazzi:2004ze} predicts an additional renormalization of the slope,
\be \alp\!_{closed} = \frac{1}{2}\left(1-\frac{c}{\lambda}\right)\alp\!_{open} \,,\ee
for some small constant \(c\), which makes this a smaller effect than that caused by the addition of the \(m_0\) mass term.

\subsection{Other string models of the glueball and the Regge slope}
In previous sections we have shown that the expected Regge slope for the closed string is

\be \alp_{closed} = \frac12\alp_{open} \,, \ee
but other string models of the glueball predict different values for the effective slope of the glueballs, \(\alp_{gb}\).

One such prediction is based on the potential between two static adjoint SU(N) charges, that, according to lattice calculations, is expected to be proportional to the quadratic Casimir operator. For small distances this added group theory factor can be obtained easily from perturbation theory, and calculations in \cite{Bali:2000un} show that what is referred to as the ``Casimir scaling hypothesis'' holds in lattice QCD for large distances as well, and this means that the effective string tension also scales like the Casimir operator (as the potential at large distances is simply \(V(\ell) \approx T_{eff}\ell\)). Therefore, a model of the glueball as two adjoint charges (or constituent gluons) joined by a flux tube predicts the ratio between the glueball and meson (two fundamental charges) slopes to be
\be \frac{\alp\!_{gb}}{\alp_{meson}} = \frac{C_2(\text{Fundamental})}{C_2(\text{Adjoint})} = \frac{N^2-1}{2N^2} = \frac{4}{9}\,, \ee
where for the last equation we take \(N = 3\). For \(N \rightarrow \infty\) we recover the ratio of \(1/2\), as can be easily seen. An argument from field theory for the double tension of the adjoint string at large \(N\) is in \cite{Armoni:2006ri}.

Other models attempt to tie the closed string to the phenomenological pomeron. The pomeron slope is measured to be \cite{Donnachie:1984xq}
\be \alp\!_{pom} = 0.25\:\text{GeV}^{-2} \approx 0.28\times\alp\!_{meson}\,, \ee
and the pomeron trajectory is commonly associated with both glueballs and closed strings. One string model that predicts a pomeron-like slope was proposed in \cite{Isgur:1984bm} and is presented in \cite{Meyer:2004jc} or in more detail in \cite{Meyer:2004gx}. It is simply the model of a rotating closed string, with a fixed circular shape. This string has two types of trajectories, a phononic trajectory (excitations propagating along the string) which has \(\alp_{phonon} = \frac{1}{4}\alp_{open}\), and an orbital trajectory (the circular string rotating around an axis in the circle's plane), for which \(\alp\!_{orbital} = \frac{3\sqrt{3}}{16}\alp_{open} \approx 0.32\times\alp_{open}\). If the rotating circular loop were allowed to deform, it would have necessarily flowed towards the flattened folded string configuration that we have been discussing, which always maximizes the angular momentum at a given energy.

There are also other possibilities of rigidly rotating closed string of other shapes, as in \cite{Burden:1982zb}, which may give yet another prediction of the ratio between open and closed string Regge slopes. Another related object is the ``\(\Delta\)-shaped'' string, which we mentioned in \cite{Sonnenschein:2014bia} as one of the stringy models of the baryon. The model is that of three masses with each pair of them connected by a string. This results in what is essentially a closed string with three quarks placed on it, which has lead 't Hooft to remark that such a configuration could be related to a quark-gluon hybrid \cite{'tHooft:2004he}, rather than a pure glueball.

\subsection{The decays of the holographic closed string} \label{sec:decays}
\subsubsection{Open string decays}
\begin{figure}[t!] \centering
	\includegraphics[width=0.95\textwidth]{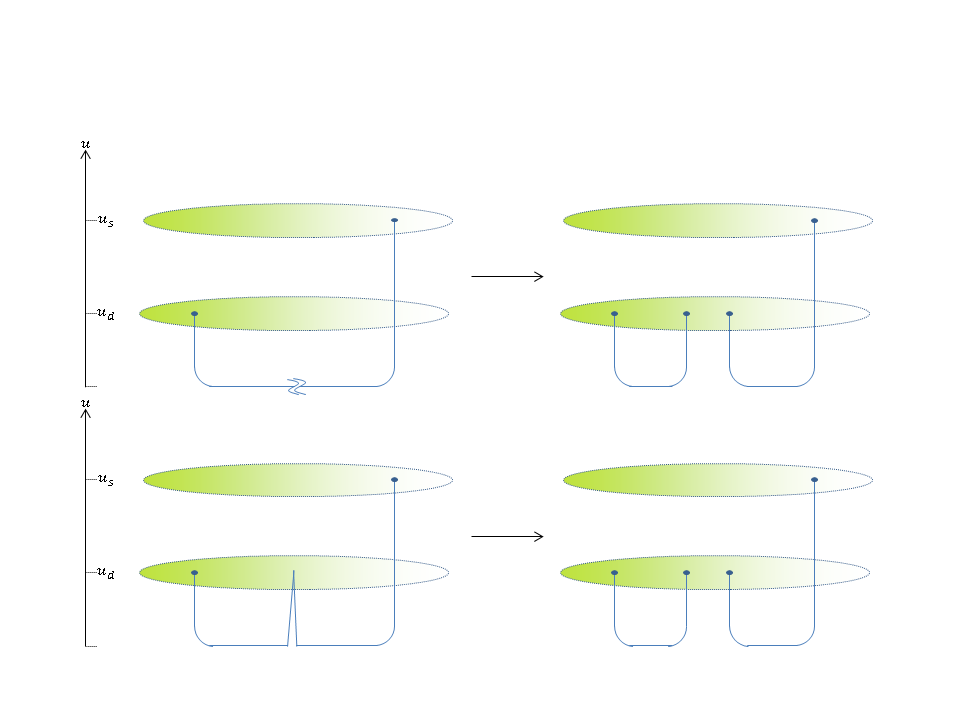}
	\caption{\label{fig:decay_open} A schematic look at the decay of a holographic open string, in this case a strange meson decaying into a strange meson and a light meson. \textbf{Top}: the picture where the string tears first, then reconnects to the flavor branes. \textbf{Bottom}: the string fluctuates up to the brane before tearing, the splits. We prefer the second picture since it assures that flavor is conserved, which is not a priori the case when the string tears at the bottom.}
			\end{figure}

The open string hadron decays when it tears at a point along the string and the two loose ends connect via quantum fluctuations to a flavor brane, creating a quark-antiquark pair. Another way to think of this process is that the string fluctuates, before tearing,	and when it reaches a flavor brane it connects to it, tears, and the pair is created. When thinking of the decay in this second way, with the fluctuation preceding the tear, it is clear that the quark and antiquark are of the same flavor, a result not a priori guaranteed when the strings tears and then reconnects to the branes. This is illustrated in figure \ref{fig:decay_open}.

The probability that a fluctuation reaches the flavor brane of a quark of flavor \(q\) is \cite{Peeters:2005fq}
\be e^{-(m_{sep}^q)^2/T} \,, \ee
where the quark mass \(m_{sep}^q\) in this context is equal to the string tension times the distance of the brane from the holographic wall.\footnote{The fact that in this model the mass \(m_{sep}^q\) is proportional to \(T\) is especially important when considering the opposing limits \(T \rightarrow 0\) and \(T\rightarrow\infty\).}

Since the tear can occur at any point along the string, we expect the total probability (and hence the total decay width) to be proportional to the string length \(L\).\footnote{In the holographic picture, it is the length of the horizontal segment of the string that is considered. When moving into flat space, it is the length between the two endpoint masses, and the relation \(M \propto TL\) receives corrections from the endpoint masses, as already written in eq. \ref{eq:holo_width} in the introduction.} We then expect that the total decay width behave like
\be \Gamma \propto Le^{-(m_{sep}^q)^2/T}\,, \ee
where \(m_{sep}^q\) is the quark produced in the decay. In \cite{Sonnenschein:2014jwa} we extracted some values of the quark masses as obtained from the Regge trajectories of mesons. For the light \(u/d\) quarks the masses were small enough so the exponent is close to one, while the \(s\) quark showed a mass for which \(m_s^2/T \sim 1\). We would then say that decays where an \(\ssb\) pair is created are suppressed by a factor of \(e^{-1}\) (before taking into account the smaller phase space).\footnote{In an alternative description \cite{Cotrone:2005fr,Bigazzi:2006jt}, the decay rate is power-like (rather than exponentially) suppressed with the mass of the quark-antiquark pair.}

\subsubsection{Rotating closed string}
\begin{figure}[tp!] \centering
	\includegraphics[width=0.95\textwidth]{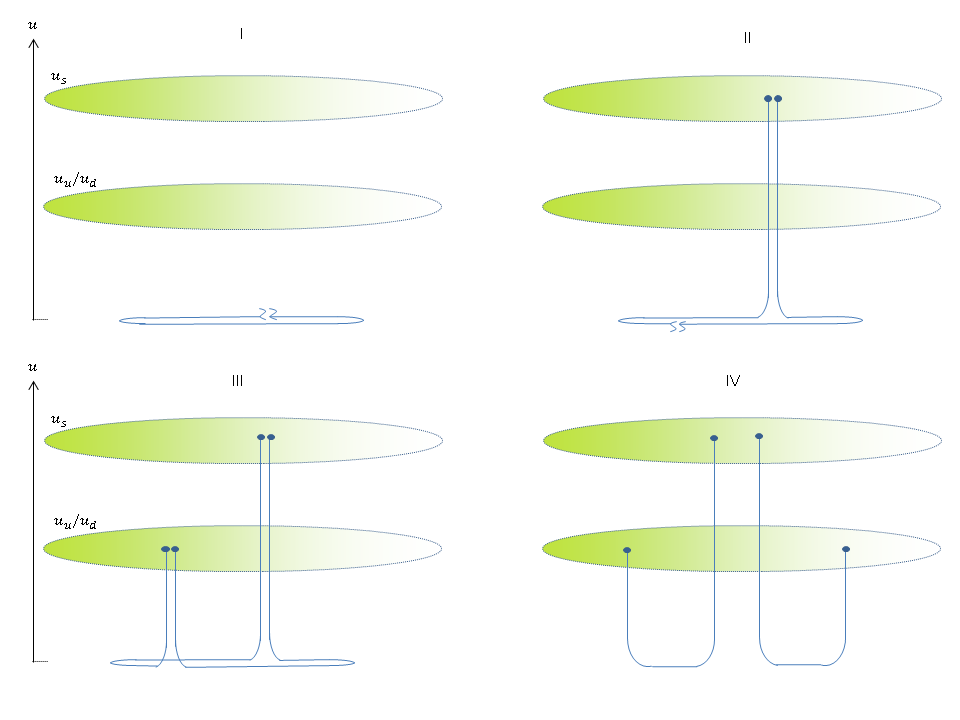} \\
	\includegraphics[width=0.67\textwidth]{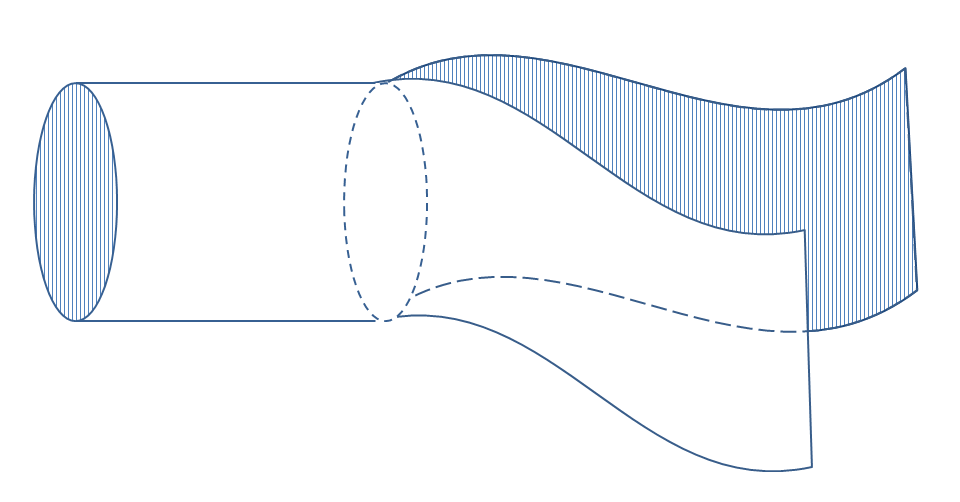}
	\caption{\label{fig:decay_closed} A schematic look at the decay of a holographic closed string to two mesons. (I) The string tears for the first time. (II) An \(\ssb\) pair is produced and the string tears for the second time. (III) A second pair is created, this time of light quarks (i.e. \(u\bar{u}\) or \(d\bar{d}\)), and two open strings are formed. (IV) A different perspective showing more clearly the final product of this decay: a \(K\) meson and a \(\bar{K}\). Note that the distances in this schematic between the flavor branes and the wall are not in scale. The bottom figure is the corresponding world sheet, of a closed string opening up at two points and forming two open strings.}
			\end{figure}

The decay process of a closed string is less simple as the string has to tear twice.\footnote{Another holographic approach to describe the decay of a glueball into two mesons, based on fields in the bulk and not closed strings was discussed in \cite{Hashimoto:2007ze,Brunner:2015oqa,Brunner:2015yha}} A single tear in the closed would produce an open string, and it in turn will have to tear again, so at the end of the process we have two open strings. If the closed string is the glueball, then this is the process of a glueball decaying to two mesons. In the total decay width we will have then the string length squared, one factor of \(L\) for each time the string tears, as well as two exponents for the two pair creation events:
\be \Gamma \propto L^2\exp(-\frac{m_q^2}{T})\exp(-\frac{m_{q^\prime}^2}{T}) \,. \label{eq:decay_closed} \ee
This process is illustrated in figure \ref{fig:decay_closed}.

If we want to identify a glueball from this basic prediction we have to look at the branching ratios of processes where the presence of the second exponent is significant, namely at processes where pairs of \(s\) and \(\bar{s}\) are produced.

The glueball unlike the meson will have the possibility of decaying into either of the three options: decay into two light mesons with two pairs of light quarks created, into \(K\bar{K}\) with one pair of \(\ssb\) and the other light, or into \(\phi\phi\) when two pairs of \(\ssb\) are created. The exponents predict the following hierarchy between the three modes:
\be \Gamma(Gb\rightarrow \text{2 light}) : \Gamma(Gb\rightarrow K\bar{K}) : \Gamma(Gb\rightarrow \phi\phi) = 1\,:e^{-1}\,:e^{-2} \,. \ee
This ratio will still need to be modified by phase space factors, which in any realistic scenario will be significant and will suppress the \(\ssb\) modes even further. This is because the states we would measure are not too far from the \(\phi\phi\) threshold of approximately 2 GeV.

\clearpage
\section{Phenomenology} \label{sec:phenomenology}
\subsection{Basic assumptions and fitting models} \label{sec:fitting_models}
We will be looking at unflavored isoscalar resonances below the \(c\bar{c}\) threshold. These states will be either mesons with the quark contents \(\frac{1}{\sqrt{2}}(u\bar{u}-d\bar{d})\) or \(s\bar{s}\), or glueballs.\footnote{Some states, such as the \(f_0(500)/\sigma\), may also be exotic multiquark states.} Correspondingly, we have several types of trajectories. For the light mesons we have the usual linear form,
\be J + n = \alp M^2 + a \,, \label{eq:traj_lin}\ee
with \(\alp = (2\pi T)^{-1}\). Note that whenever we use \(\alp\) without a subscript in this paper, it refers to this slope of the linear meson trajectories.

For \(s\bar{s}\) states, we use the formula for the mass corrected trajectory (as was used in \cite{Sonnenschein:2014jwa}) defined by
	\be E = 2m_s\left(\frac{\beta\arcsin \beta+\sqrt{1-\beta^2}}{1-\beta^2}\right) \label{eq:massFitE} \,,\ee
	\be J + n = a + 2\pi\alp m_s^2\frac{\beta^2}{(1-\beta^2)^2}\left(\arcsin \beta+\beta\sqrt{1-\beta^2}\right) \,.\label{eq:traj_mass} \ee
These are the trajectories of a rotating string with two masses \(m_s\) at its endpoints, and with an added intercept and extrapolated \(n\) dependence. \(\beta\) is the velocity of the endpoint mass. The limit \(m_s \rightarrow 0\) (with \(\beta \rightarrow 1\)) takes us back to the linear trajectory of eq. \ref{eq:traj_lin}, with the first correction in the expression for \(J\) being proportional to \(\alp m_s^{3/2} E^{1/2}\).

For the glueballs we assume linear trajectories of the form
\be J + n = \alp\!_{gb} M^2 + a \,,\ee
and we take \(\alp\!_{gb}\) to be \(\frac{1}{2}\alp\), where \(\alp\) is the slope of the mesons as obtained in our fits of the various meson trajectories. A typical value would be between 0.80 and 0.90 GeV\(^{-2}\).

In a later section we examine the possible application of the formula based on the holographic prediction,
\be J + n = \alp_{gb}E^2-2\alp_{gb}m_0E + a\,.\ee
When using this formula we will also take \(\alp\!_{gb} = \frac{1}{2}\alp\), ignoring the possible correction to the slope, which we assume to be small.

One assumption which we must state explicitly before continuing to the fits is that there is no mixing of light mesons, \(\ssb\) mesons, and glueballs. It is an open question how strongly glueballs and mesons are mixed, with results varying greatly between different models, from almost maximal mixing to very weak (different results based on different models are collected in \cite{Crede:2008vw}). In a stringy model, where glueballs are represented by closed strings and mesons by open strings, it seems more natural that they will not mix at all. We also assume that the mixing between the light quark states and the \(\ssb\) is weak, in placing states either on the linear trajectories of the light mesons or on the mass corrected trajectories of the \(\ssb\), the same assumption that was used in \cite{Sonnenschein:2014jwa} in fitting the \(\omega\) and \(\phi\) mesons. It is not obvious how the possible mixing between the two types of mesons affects the trajectories.

\subsubsection{The two types of trajectories}
Along \emph{radial trajectories}, or trajectories in the \((n,M^2)\) plane, the states differ only by the radial\footnote{The term should not be confused as having something to do with the radial coordinate of holography.} excitation number \(n\), all other quantum numbers constant. Since \(n\) is not actually measured we have to assign a value ourselves to the different states, and from there emerges a great ambiguity that we have to solve.

Mesons belong on trajectories in the \((n,M^2)\) plane with a slope that seems to be slightly smaller than in the \((J,M^2)\) plane. The typical values are 0.80--0.85 GeV\(^{-2}\) for the former and \(0.90\) GeV\(^{-2}\) for the latter type of trajectories, as our fits in \cite{Sonnenschein:2014jwa} have shown. We implicitly assume in the following sections that for the glueballs there will be a similar difference between the slopes in the different planes. When we write that \(\alp_{gb} = \frac12\alp_{meson}\) we refer to \(\alp_{meson}\) as it is obtained for the meson fits in the same plane, rather than taking fixed values of \(\alp\). This also serves to restrict the number of parameters in a given fit: we always try to describe all the trajectories using a single value of \(\alp\).

We should also note that while for the mesons, \(n\) naturally takes the values \(n = 0,1,2,\ldots\) along the radial trajectories, the case is not so for glueballs. For the closed strings we noted in section \ref{sec:lightcone} that the number of left and right moving modes has to be equal, and so \(n\), which is really \(N+\tilde{N}\) in this case, should be even: \(n = 0,2,4,\ldots\).

For the \emph{orbital trajectories}, or trajectories in the \((J,M^2)\) plane, we expect to find, along the leading trajectory of the glueball, the ground state with \(J^{PC} = 0^{++}\) followed by the tensor glueball (\(2^{++}\)) as its first excited state, and continue to higher states with even \(J\) and \(PC = ++\).

The orbital trajectories of the mesons will be constructed as usual, and using the known quark model relations \(P = (-1)^{L+1}\) and \(C = (-1)^{L+S}\). The relevant trajectories are then expected to have states with \(J^{PC} = 1^{--}, 2^{++}, 3^{--}, 4^{++}, \ldots\). It is worth noting then that for mesons, a \(0^{++}\) state is an excited state with \(L = 1\) and \(S = 1\), and not a part of what we usually take for the trajectory when we use states of increasing \(J\).

\subsection{The glueball candidates: The \texorpdfstring{$f_0$}{f0} and \texorpdfstring{$f_2$}{f2} resonances}
There is an abundance of isoscalar states with the quantum numbers \(J^{PC} = 0^{++}\) (the \(f_0\) resonances) or \(J^{PC} = 2^{++}\) (\(f_2\)). The Particle Data Group's (PDG) latest Review of Particle Physics \cite{PDG:2014}, which we we use as the source of experimental data throughout this paper, lists 9 \(f_0\) states and 12 \(f_2\) states, with an additional 3 \(f_0\)'s and 5 \(f_2\)'s listed as unconfirmed ``further states''. These are listed in tables \ref{tab:allf0} and \ref{tab:allf2}. In the following we make a naive attempt to organize the known \(f_0\) and \(f_2\) states into trajectories, first in the plane of orbital excitations \((J,M^2)\), then in the radial excitations plane \((n,M^2)\).

The states classified as ``further states'' are generally not used unless the prove to be necessary to complete the trajectories formed by the other states. The ``further states'' will be marked with an asterisk below.\footnote{Note that the asterisk is not standard notation nor a part of the PDG given name of a state, we only use it to make clear the status of given states throughout the text.}

It is not the purpose of this paper to review all the information available on the \(f_0\) and \(f_2\) resonances, nor to present the different theories and speculations regarding their meson or glueball nature. We usually attempt to form Regge trajectories first, using just the masses and basic quantum numbers, and then verify if the implications regarding the contents of a given state make sense in the light of additional experimental data, namely the different states' decay modes.

For a more complete picture regarding the spectrum and specifically the interpretation of the different resonances as glueballs, the reader is referred to reviews on glueball physics and their experimental status such as \cite{Klempt:2007cp,Mathieu:2008me,Crede:2008vw,Ochs:2013gi}, citations therein, and subsequent works citing these reviews.

\begin{table}[t!] \centering
		\begin{tabular}{|l|l|l|l|l|} \hline
		\textbf{State} & \textbf{Mass} [MeV] & \textbf{Width} [MeV] & \textbf{Width/mass} & \textbf{Decay modes} \\ \hline\hline
		\(f_0(500)/\sigma\) & 400--550 & 400--700 & 1.16\plm0.36 & \(\pi\pi\) dominant\\ \hline
		\(f_0(980)\) & \(990\pm20\) & 40--100 & 0.07\plm0.03 & \(\pi\pi\) dominant, \(K\overline{K}\) seen \\ \hline
		\(f_0(1370)\) & 1200--1500 & 200--500 & 0.26\plm0.11 & \(\pi\pi\), \(4\pi\), \(\eta\eta\), \(K\overline{K}\) \\ \hline
		\(f_0(1500)\) & \(1505\pm6\) & \(109\pm7\) & 0.072\plm0.005 & \(\pi\pi\) \([35\%]\), \(4\pi\) \([50\%]\), \\
									&  & & & \(\eta\eta\)/\(\eta\eta\prime\) \([7\%]\), \(K\overline{K}\) \([9\%]\) \\ \hline		
		\(f_0(1710)\) & \(1720\pm6\) & \(135\pm8\) & 0.078\plm0.005 & \(K\overline{K}\), \(\eta\eta\), \(\pi\pi\) \\ \hline
		\(f_0(2020)\) & \(1992\pm16\) & \(442\pm60\) & 0.22\plm0.03 & \(\rho\pi\pi\), \(\pi\pi\), \(\rho\rho\), \(\omega\omega\), \(\eta\eta\) \\ \hline
		\(f_0(2100)\) & \(2103\pm8\) & \(209\pm19\) & 0.10\plm0.01 & \\ \hline
		\(f_0(2200)\) & \(2189\pm13\) & \(238\pm50\) & 0.11\plm0.02 & \\ \hline
		\(f_0(2330)\) & \(2325\pm35\) & \(180\pm70\) & 0.08\plm0.03 &  \\ \hline
		*\(f_0\)(1200--1600) & 1200--1600 & 200--1000 & 0.43\plm0.29 & \\ \hline
		*\(f_0\)(1800) & \(1795\pm25\) & \(95\pm80\) & 0.05\plm0.04 & \\ \hline
		*\(f_0\)(2060) & \(\sim2050\) & \(\sim120\) & \(\sim0.04\)--\(0.10\) & \\ \hline
		\end{tabular} \caption{\label{tab:allf0} All the \(f_0\) states as listed by the PDG. The last few states, marked here by asterisk, are classified as ``further states''.}
	\end{table}

		\begin{table}[t!] \centering
		\begin{tabular}{|l|l|l|l|l|} \hline
		\textbf{State} & \textbf{Mass} [MeV] & \textbf{Width} [MeV] & Width/mass & \textbf{Decay modes} \\ \hline\hline
		\(f_2(1270)\) & 1275.1\plm1.2 & 185.1\plm2.9 & 0.15\plm0.00 & \(\pi\pi\) \([85\%]\), \(4\pi\) \([10\%]\), \(KK\), \(\eta\eta\), \(\gamma\gamma\), ... \\ \hline
		\(f_2(1430)\) & 1453\plm4 & 13\plm5 & 0.009\plm0.006 & \(KK\), \(\pi\pi\) \\ \hline
		\(f^\prime_2(1525)\) & 1525\plm5 & 73\plm6 & 0.048\plm0.004 & \(KK\) \([89\%]\), \(\eta\eta\) \([10\%]\), \(\gamma\gamma\) [seen], ... \\ \hline
		\(f_2(1565)\) & 1562\plm13 & 134\plm8 & 0.09\plm0.01 & \(\pi\pi\), \(\rho\rho\), \(4\pi\), \(\eta\eta\), ... \\ \hline
		\(f_2(1640)\) & 1639\plm6 & 99\plm60 & 0.06\plm0.04 & \(\omega\omega\), \(4\pi\), \(KK\) \\ \hline
		\(f_2(1810)\) & 1815\plm12 & 197\plm22 & 0.11\plm0.01 & \(\pi\pi\), \(\eta\eta\), \(4\pi\), \(KK\), \(\gamma\gamma\) [seen] \\ \hline
		\(f_2(1910)\) & 1903\plm9 & 196\plm31 & 0.10\plm0.02 & \(\pi\pi\), \(KK\), \(\eta\eta\), \(\omega\omega\), ... \\ \hline
		\(f_2(1950)\) & 1944\plm12 & 472\plm18 & 0.24\plm0.01 & \(K^*K^*\), \(\pi\pi\), \(4\pi\), \(\eta\eta\), \(KK\), \(\gamma\gamma\), \(pp\) \\ \hline
		\(f_2(2010)\) & 2011\plm76 & 202\plm67 & 0.10\plm0.03 & \(KK\), \(\phi\phi\) \\ \hline
		\(f_2(2150)\) & 2157\plm12 & 152\plm30 & 0.07\plm0.01 & \(\pi\pi\), \(\eta\eta\), \(KK\), \(f_2(1270)\eta\), \(a_2\pi\), \(pp\) \\ \hline
		\(f_J(2220)\) & 2231.1\plm3.5 & 23\plm8 & 0.010\plm0.004 & \(\pi\pi\), \(KK\), \(pp\), \(\eta\eta^\prime\) \\ \hline
		\(f_2(2300)\) & 2297\plm28 & 149\plm41 & 0.07\plm0.02 & \(\phi\phi\), \(KK\), \(\gamma\gamma\) [seen] \\ \hline
		\(f_2(2340)\) & 2339\plm55 & 319\plm81 & 0.14\plm0.04 & \(\phi\phi\), \(\eta\eta\) \\ \hline
		*\(f_2(1750)\) & 1755\plm10 & 67\plm12 & 0.04\plm0.01 & \(KK\), \(\gamma\gamma\), \(\pi\pi\), \(\eta\eta\) \\ \hline
		*\(f_2(2000)\) & 2001\plm10 & 312\plm32 & 0.16\plm0.02 & \\ \hline
		*\(f_2(2140)\) & 2141\plm12 & 49\plm28 & 0.02\plm0.01 & \\ \hline
		*\(f_2(2240)\) & 2240\plm15 & 241\plm30 & 0.11\plm0.01 & \\ \hline
		*\(f_2(2295)\) & 2293\plm13 & 216\plm37 & 0.10\plm0.02 & \\ \hline
		\end{tabular} \caption{\label{tab:allf2} All the \(f_2\) states as listed by the PDG. The last few states, marked here by asterisk, are classified as ``further states''.}
	\end{table}

	\subsection{Assignment of the \texorpdfstring{$f_0$}{f0} into trajectories} \label{sec:f0_fits}
	In a given assignment, we generally attempt to include all the \(f_0\) states listed in table \ref{tab:allf0}, sorting them into meson and, if possible, glueball trajectories.
	
	We make an exception of the \(f_0(500)/\sigma\) resonance, which we do not use in any of the following sections. Its low mass and very large width are enough to make it stand out among the other \(f_0\) states listed in the table. There is no common consensus regarding the composition of the \(\sigma\). We find that it does not belong on a meson Regge trajectory. If we assume it is a glueball then our model predicts the next state to be at around 2.2 GeV, and, since we assume its width to be proportional to its mass squared (as implied by eq. \ref{eq:decay_closed}), it would have a width of at least 8 GeV. We hope, in that case, that there is no reason to make such an assumption.\footnote{The authors of \cite{Nebreda:2011cp} state that the interpretation of the \(f_0(500)/\sigma\) as a glueball is ``strongly disfavored'', from what they consider a model independent viewpoint. We found no references that suggest the opposite.} Therefore, we simply ``ignore'' the \(f_0(500)\) in the following sections.

	\subsubsection{Assignment of all states as mesons}
		Sorting the \(f_0\) states into trajectories with a meson-like slope leads to an assignment of the \(f_0\)'s into two groups of four:
	\[\mathrm{Light}:\qquad980, 1500, 2020, 2200, \]
	\[\ssb:\qquad1370, 1710, 2100, 2330. \]
While this simple assignment includes all the confirmed \(f_0\) states (except the \(f_0(500)\)) on two parallel trajectories, it remains unsatisfactory. If there are no glueballs we expect the states in the lower trajectory to be (predominantly) composed of light quarks, while the higher states should be \(\ssb\). This does not match what we know about the decay modes of the different states. For example, the \(f_0(1370)\) does not decay nearly as often to \(K\bar{K}\) as one would expect from an \(\ssb\) state. In fact, this assignment of the \(f_0\)'s into meson trajectories was proposed in some other works \cite{Anisovich:2000kxa,Anisovich:2002us,Masjuan:2012gc}, and the mismatch with the decay modes was already addressed in greater detail in \cite{Bugg:2012yt}.

	\subsubsection{Assignment with \texorpdfstring{$f_0$}{f0}(980) as glueball}
	\begin{table}[tp!] \centering
	\includegraphics[width=0.60\textwidth]{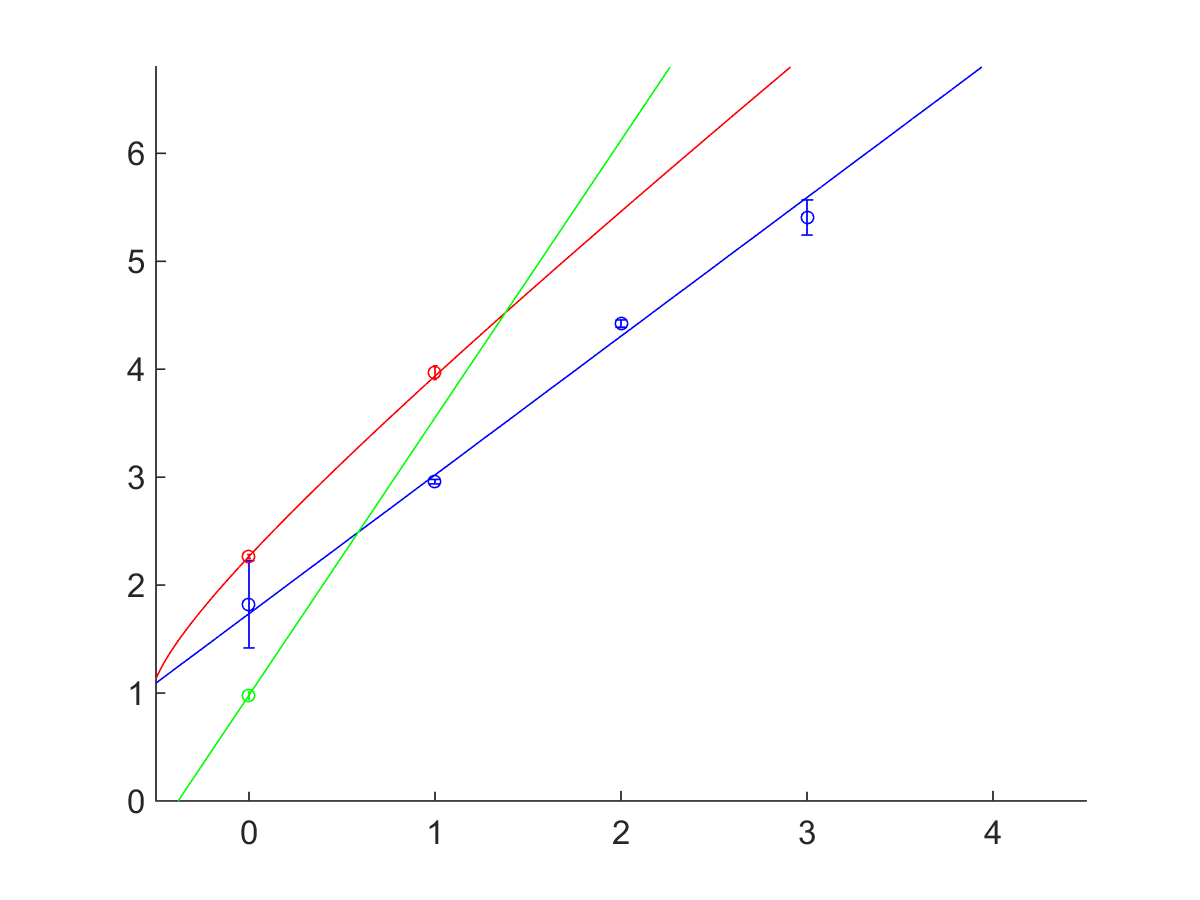} \qquad
				\begin{tabular}{|c|c|c|c|c|c|c|} \hline
					\(n\) & \multicolumn{2}{|c|}{Light} & \multicolumn{2}{|c|}{\(s\bar{s}\)} & \multicolumn{2}{|c|}{Glueball} \\ \hline
					& Exp. & Thry. & Exp. & Thry. & Exp. & Thry. \\ \hline
			0 &	1350\plm150 & 1317 & 1505\plm6 & 1505 & 990\plm20 & 990 \\
			1 &	1720\plm6 & 1738 & 1992\plm16 & 1984 & - & - \\
			2 &	2103\plm8 & 2075 & ? & 2340 & ? & 2470 \\
			3 &	2325\plm35 & 2365 & & & & \\
			4 & ? & 2620 & & & & \\ \hline
				\end{tabular} \caption{\label{tab:fit980} The results of the fit to the assignment with \(f_0(980)\) as the glueball ground state. The slope is \(\alp = 0.788\) GeV\(^{-2}\) and the mass of the \(s\) quark \(m_s = 500\) MeV. This fit has \(\chi^2 = 3.78\). The intercepts obtained are (-1.35) for light mesons, (-0.52) for \(\ssb\), and (-0.38) for glueballs. We also list the predicted mass of the next state in each trajectory.}
			\end{table}

	In this and the following sections we pick and single out a state as the glueball ground state and try to build the meson trajectories without it.

	First is the the \(f_0(980)\). Assuming it is the glueball then the \(f_0(2330)\) is at the right mass to be its first excited (\(n = 2\)) partner. However, we find that the two meson trajectories given this assignment,
\[\mathrm{Light:}\qquad 1370, 1710, 2100,\]
\[\ssb\mathrm{:}\qquad 1500, 2020,\]
also predict a state very near the mass of the \(f_0(2330)\), and according to this assignment, there should be two more \(f_0\) states near the \(f_0(2330)\), for a total of three. The \(f_0(2200)\) has to be excluded.

We again have to put some states on trajectories that are not quite right for them: the \(f_0(1710)\) has a significant branching ratio for its decay into \(K\overline{K}\), while the \(f_0(1500)\), which is taken as the head of the \(\ssb\) trajectory, decays to \(K\overline{K}\) less than \(10\%\) of the time.

Note that the assignment above is the same as the one we would make if we excluded the \(f_0(980)\) on the grounds of it being an exotic (but non-glueball) state and assumed all the other states are mesons. The \(f_0(980)\) is commonly believed to be a multiquark state or a \(K\bar{K}\) ground state,\footnote{See the PDG's ``Note on scalar mesons below 2 GeV'' and references therein.} and in fact, we will find in following sections that even it is not a glueball, it is better to exclude it from the meson trajectories. The trajectories and masses obtained are in table \ref{tab:fit980}.
	
	\subsubsection{Assignment with \texorpdfstring{$f_0$}{f0}(1370) as glueball}
	\begin{table}[tp!] \centering
	\includegraphics[width=0.60\textwidth]{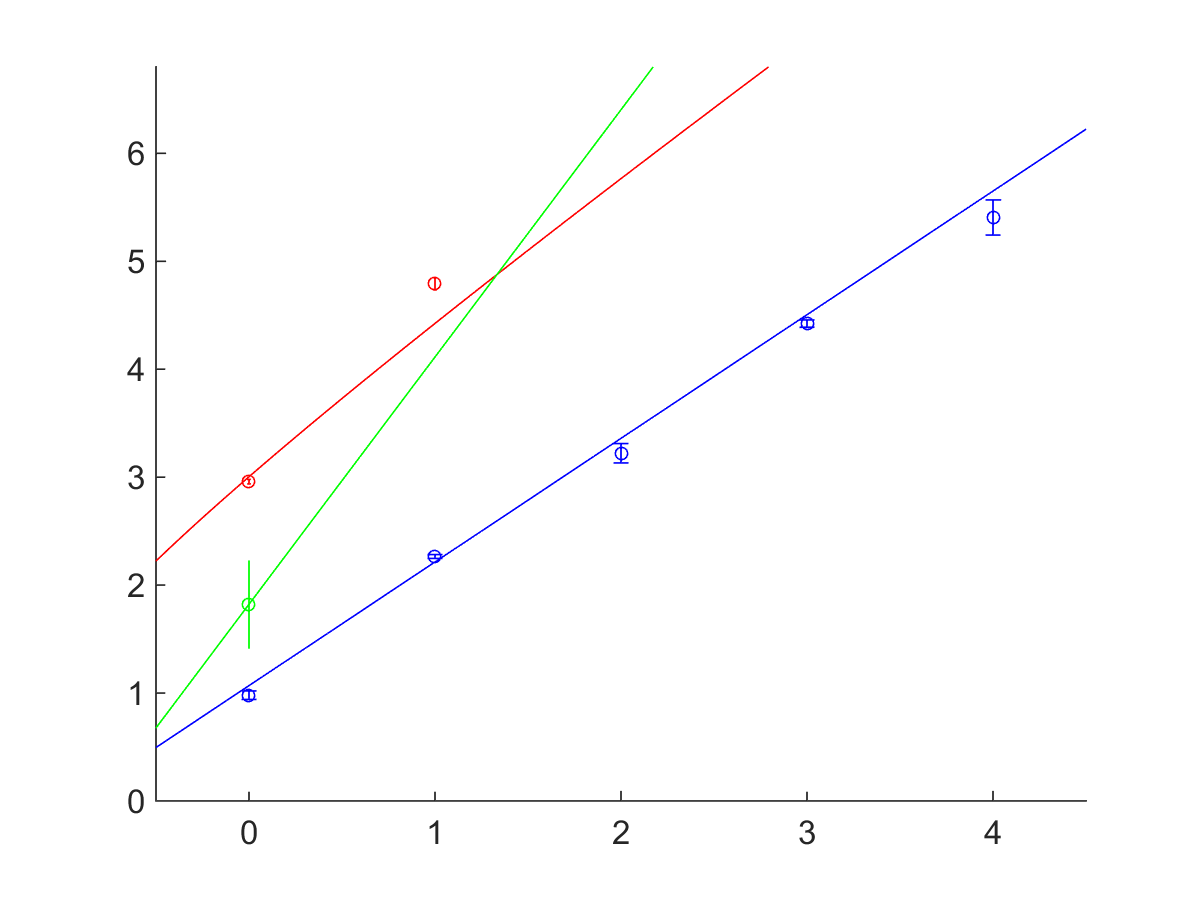} \qquad
				\begin{tabular}{|c|c|c|c|c|c|c|} \hline
					\(n\) & \multicolumn{2}{|c|}{Light} & \multicolumn{2}{|c|}{\(s\bar{s}\)} & \multicolumn{2}{|c|}{Glueball} \\ \hline
					& Exp. & Thry. & Exp. & Thry. & Exp. & Thry. \\ \hline
			0 &	990\plm20 & 1031 & 1720\plm6 & 1733 & 1350\plm150 & 1350 \\
			1 &	1505\plm6 & 1488 & 2189\plm13 & 2103 & - & - \\
			2 &	1795\plm25 & 1835 & ? & 2400 & ? & 2530 \\
			3 &	2103\plm8 & 2123 & & & & \\
			4 & 2325\plm35 & 2377 & & & & \\
			5 & ? & 2610 & & & & \\ \hline
				\end{tabular} \caption{\label{tab:fit1370} The results of the fit to the assignment with \(f_0(1370)\) as the glueball ground state. The slope is \(\alp = 0.873\) GeV\(^{-2}\) and the mass of the \(s\) quark \(m_s = 500\) MeV. This fit has \(\chi^2 = 10.01\). The intercepts obtained are (-0.93) for light mesons, (-1.06) for \(\ssb\), and (-0.80) for glueballs. We also list the predicted mass of the next state in each trajectory.}
			\end{table}

From here onwards the states singled out as glueballs are too high in mass for their excited states to be in the range of the \(f_0\) states listed in table \ref{tab:allf0}, that is beneath 2.4 GeV.

Excluding the \(f_0(1370)\), we have:
\[\mathrm{Light:}\qquad [980], 1500, *1800, 2100, 2330\]
\[\ssb\mathrm{:}\qquad 1710, 2200.\]
The \(f_0(980)\) is put here in brackets to emphasize that it is optional. Including or excluding it can affect some of the fitting parameters but the trajectory is certainly not incomplete if we treat \(f_0(980)\) as a non-meson resonance and take \(f_0(1500)\) as the head of the trajectory.

The main issue here is that we have to use the state \(*f_0(1800)\) to fill in a hole in the meson trajectory, a state that is still considered unconfirmed by the PDG and whose nature is not entirely known. It was observed so far only as an enhancement in the radiative decay \(J/\psi \rightarrow \gamma\omega\phi\) and its observers at BESIII \cite{Ablikim:2012ft} suggest it is an exotic state - a tetraquark, a hybrid, or itself a glueball. More experimental data is needed here.

Other than that we have \(f_0(2100)\) as a light meson and \(f_0(2200)\) as \(\ssb\). This is the option that is more consistent with the decays, as \(f_0(2200)\) is the one state of the two which is known to decay into \(K\overline{K}\) (we again refer to the comments in \cite{Bugg:2012yt} and references therein). However, in terms of the fit, we might do better to exchange them. It is possible that the proximity of these two resonances to each other affects their masses in such a way that our model can not predict, and this affects badly the goodness of our fit, as can be seen in table \ref{tab:fit1370}.

\subsubsection{Assignment with \texorpdfstring{$f_0$}{f0}(1500) as glueball}

\begin{table}[tp!] \centering
	\includegraphics[width=0.60\textwidth]{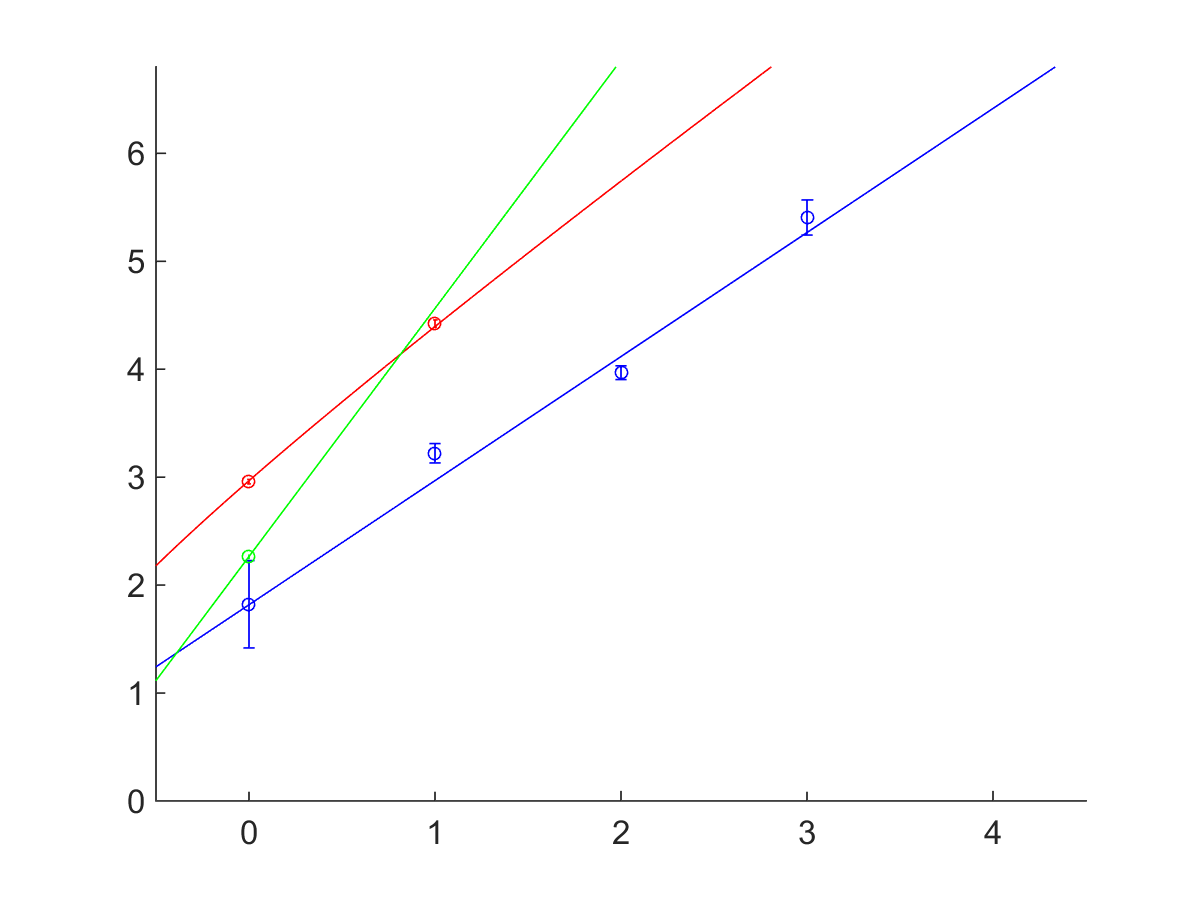} \qquad
				\begin{tabular}{|c|c|c|c|c|c|c|} \hline
					\(n\) & \multicolumn{2}{|c|}{Light} & \multicolumn{2}{|c|}{\(s\bar{s}\)} & \multicolumn{2}{|c|}{Glueball} \\ \hline
					& Exp. & Thry. & Exp. & Thry. & Exp. & Thry. \\ \hline
			0 &	1350\plm150 & 1031 & 1720\plm6 & 1723 & 1505\plm6 & 1505 \\
			1 &	1795\plm25 & 1723 & 2103\plm8 & 2097 & - & - \\
			2 &	1992\plm16 & 2029 & ? & 2400 & ? & 2620 \\
			3 &	2325\plm35 & 2295 & & & & \\
			4 & ? & 2530 & & & & \\ \hline
				\end{tabular} \caption{\label{tab:fit1500} The results of the fit to the assignment with \(f_0(1500)\) as the glueball ground state. The slope is \(\alp = 0.870\) GeV\(^{-2}\) and the mass of the \(s\) quark \(m_s = 500\) MeV. This fit has \(\chi^2 = 2.51\). The intercepts obtained are (-1.58) for light mesons, (-1.03) for \(\ssb\), and (-0.99) for glueballs. We also list the predicted mass of the next state in each trajectory.}
			\end{table}
			
Taking the \(f_0(1500)\) to be the glueball, then the light meson trajectory will start with \(f_0(1370)\), giving:
	\[\mathrm{Light:}\qquad 1370, *1800, 2020, 2330,\]
	\[\ssb\mathrm{:}\qquad 1710, 2100.\]
With \(f_0(1500)\) identified as the glueball, this assignment includes all the states except \(f_0(2200)\). Incidentally though, the \(f_0(2200)\) would have belonged on the glueball trajectory if we had allowed odd values of \(n\) for the glueball. In other words, it matches the prediction for the \(n = 1\) state of the half slope trajectory beginning with \(f_0(1500)\). We could also use \(f_0(2200)\) as the \(\ssb\) state and leave out \(f_0(2100)\) instead.

There is no glaring inconsistency in this assignment with the decay modes, but we are again confronted with the state \(*f_0(1800)\), which we need to complete the light meson trajectory. We can see from table \ref{tab:allf0} that the \(f_0(2020)\) is wider than other states in its trajectory, whereas we maintain that the ratio between width and mass \(\Gamma/M\) should be roughly constant along a trajectory. In particular, the last state in the trajectory, \(f_0(2330)\), is much narrower than \(f_0(2020)\). We can assign the \(f_0(2330)\) to the \(\ssb\) trajectory instead, but there is no other argument for that state being \(\ssb\), considering it was observed only in its decays to \(\pi\pi\) and \(\eta\eta\). Perhaps the fact that \(f_0(1370)\) and \(f_0(2020)\) are both quite wide means that there should be two additional states, with masses comparable to those of \(*f_0(1800)\) and \(f_0(2330)\), that are also wide themselves, and those states will better complete this assignment. The results of the assignment are presented in table \ref{tab:fit1500}.
			
\subsubsection{Assignment with \texorpdfstring{$f_0$}{f0}(1710) as glueball}

\begin{table}[tp!] \centering
	\includegraphics[width=0.60\textwidth]{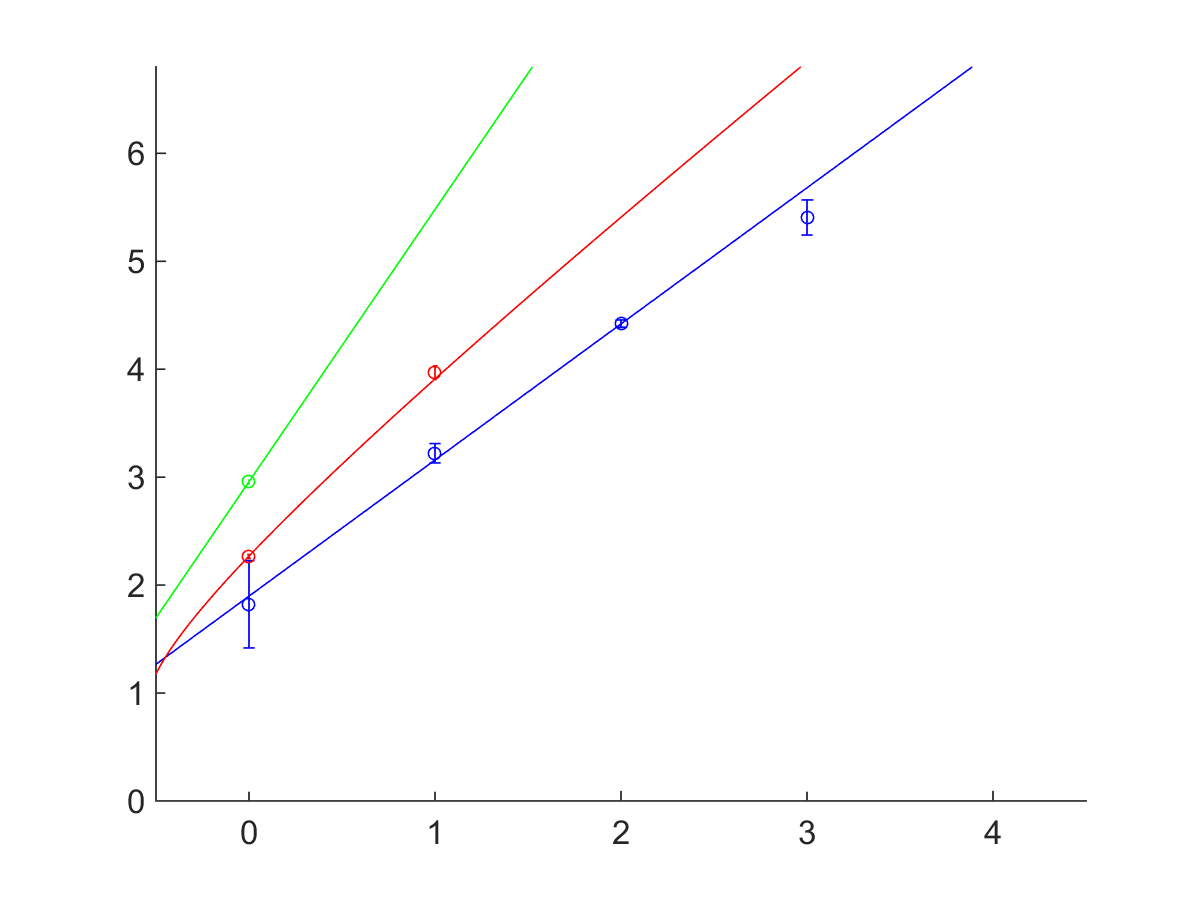} \qquad
				\begin{tabular}{|c|c|c|c|c|c|c|} \hline
					\(n\) & \multicolumn{2}{|c|}{Light} & \multicolumn{2}{|c|}{\(s\bar{s}\)} & \multicolumn{2}{|c|}{Glueball} \\ \hline
					& Exp. & Thry. & Exp. & Thry. & Exp. & Thry. \\ \hline
			0&	1350\plm150 & 1378 & 1505\plm6 & 1506 & 1720\plm6 & 1720 \\
			1&	1795\plm25 & 1777 & 1992\plm16 & 1977 & - & - \\
			2&	2103\plm8 & 2102 & ? & 2330 & ? & 2830 \\
			3&	2325\plm35 & 2383 & & & & \\
			4&	? & 2640 & & & & \\ \hline
				\end{tabular} \caption{\label{tab:fit1710} The results of the fit to the assignment with \(f_0(1710)\) as the glueball ground state. The slope is \(\alp = 0.793\) GeV\(^{-2}\) and the mass of the \(s\) quark \(m_s = 500\) MeV. This fit has \(\chi^2 = 0.71\). The intercepts obtained are (-1.51) for light mesons, (-0.53) for \(\ssb\), and (-1.17) for glueballs.}
			\end{table}
			
Excluding the \(f_0(1710)\) from the meson trajectories we can make an assignment that includes all states except the \(f_0(500)\) and \(f_0(980)\):
\[\mathrm{Light:}\qquad 1370, *1800, 2100, 2330\]
\[\ssb\mathrm{:}\qquad 1500, 2020, 2200\]
\[\mathrm{Glue:}\qquad 1710\]
The disadvantage here is that we again have to use \(f_0(1500)\) as the head of the \(\ssb\) trajectory despite knowing that its main decay modes are to \(4\pi\) and \(\pi\pi\), as well as the fact the we - once again - need the \(*f_0(1800)\) resonance to fill in a hole for \(n = 1\) in the resulting light meson trajectory. This trajectory can be seen in table \ref{tab:fit1710}.

\subsubsection{Conclusions from the \texorpdfstring{$f_0$}{f0} fits}
It is not hard to see that the \(f_0\) resonances listed in the PDG's Review of Particle Physics all fit in quite neatly on two parallel trajectories with a slope similar to that of other mesons. However, upon closer inspection, these trajectories - one for light quark mesons and one for \(\ssb\) - are not consistent with experimental data, as detailed above. For us the naive assignment is also inconsistent with what we have observed for the other \(\ssb\) trajectories in \cite{Sonnenschein:2014jwa}, namely that the \(\ssb\) trajectories are not purely linear, and have to be corrected by adding a non-zero string endpoint mass for the \(s\) quark, usually of at least 200 MeV.

The other novelty that we hoped to introduce, the half slope trajectories of the glueball, proved to be impractical - given the current experimental data which only goes up to less than \(2.4\) GeV for the relevant resonances.

	One conclusion that can be drawn is that the state \(f_0(980)\) can be comfortably excluded from any of the meson trajectories, which is consistent with its being the \(K\overline{K}\) ground state.

	The unconfirmed state \(*f_0(1800)\) turns up in the assignments with glueballs in them, usually to fill in a hole in the light meson trajectory. If the \(*f_0(1800)\) is not in itself a meson as mentioned before, then we would hope that there is another yet unobserved \(f_0\) state with a very similar mass, say 1800--1850 MeV.
	
	There is no one assignment that seems the correct one, although the two assignments singling out either \(f_0(1370)\) or \(f_0(1500)\) as the glueball ground states seem more consistent than the other possibilities. The best way to determine which is better is, as always, by finding more experimental data. We list our predictions for higher resonances based on these assignments in section \ref{sec:predictions} of the appendix.

\subsection{Assignment of the \texorpdfstring{$f_2$}{f2} into trajectories} \label{sec:f2_fits}
We now turn to the \(f_2\) tensor resonances, that were listed in the beginning of the section in table \ref{tab:allf2}. We will first examine trajectories in the \((J,M^2)\) plane, then move on to the attempt to assign all the \(f_2\) states to trajectories in the \((n,M^2)\) plane.
\subsubsection{Trajectories in the \texorpdfstring{$(J,M^2)$}{(J,M2)} plane} \label{sec:f2_orbital}
The only way to get a linear trajectory connecting a \(0^{++}\) and a \(2^{++}\) state with the slope \(\alp\!_{gb} = \frac{1}{2}\alp_{meson}\) is to take the lightest \(f_0\) glueball candidate and the heaviest known \(f_2\). Then we have the pair \(f_0(980)\) and \(f_2(2340)\), and the straight line between them has a slope of 0.45 GeV\(^{-2}\). There is no \(J = 1\) resonance near the line stretched between them. However, this example mostly serves to demonstrate once again the difficulty of forming the glueball trajectories in practice. The glueball states are predicted to be fewer and farther apart then the mesons in their respective Regge trajectories.

Therefore, it is a more sound strategy to look again for the meson trajectories, see what states are excepted from them, and check for overall consistency of the results. In forming the meson trajectories, we know that we can expect the \(\omega\) mesons with \(J^{PC} = 1^{--}\) to be part of the trajectories, in addition to some states at higher spin, which will allow us to form trajectories with more points.

Moving on from \(J = 0^{++}\) and \(2^{++}\) to higher spin states, we see two \(J^{PC} = 4^{++}\) states that could belong to a trajectory: \(f_4(2050)\) and \(f_4(2300)\). The first of those, \(f_4(2050)\), belongs to a well known meson trajectory in the \((J,M^2)\) plane, following \(\omega(782)\), \(f_2(1270)\), and \(\omega_3(1670)\). The slope of the fit to that trajectory is \(\alp = 0.91\) GeV\(^{-2}\), and we can even include in it states of spin 5 and 6: \(*\omega_5(2250)\) and \(f_6(2510)\).

The mass of the \(f_4(2300)\) is too low for it to belong to a linear trajectory with a glueball slope. Taking it to be a meson one can put it on a linear trajectory following \(\omega(1420)\) and \(f_2(1810)\). To complete this trajectory we need a \(J^{PC} = 3^{--}\) state with a mass near 2070 MeV. The PDG lists one unconfirmed state, \(X(2080)\), with the quantum numbers \(I(J^{PC}) = ?(3^{-?})\), which might be a match.

We also find another meson trajectory involving the second excited \(\omega\) meson - \(\omega(1650)\). This trajectory would be comprised of \(\omega(1650)\), \(*\omega_3(2255)\), and with one of \(f_2(1950)\) or \(f_2(2010)\) between them.

We also have the meson trajectories of the \(\ssb\). The first joins the ground state \(\phi(1020)\) with \(f_2^\prime(1525)\) and \(\phi_3(1850)\). We can form a daughter trajectory starting with the \(\phi(1680)\), and going on to include \(f_2(1950)\) or \(f_2(2010)\), as well as the unconfirmed \(*\omega_3(2285)\). This trajectory is nearly identical to that of the \(\omega(1650)\) of the last paragraph.

\begin{figure}[t!] \centering
	\includegraphics[width=0.76\textwidth]{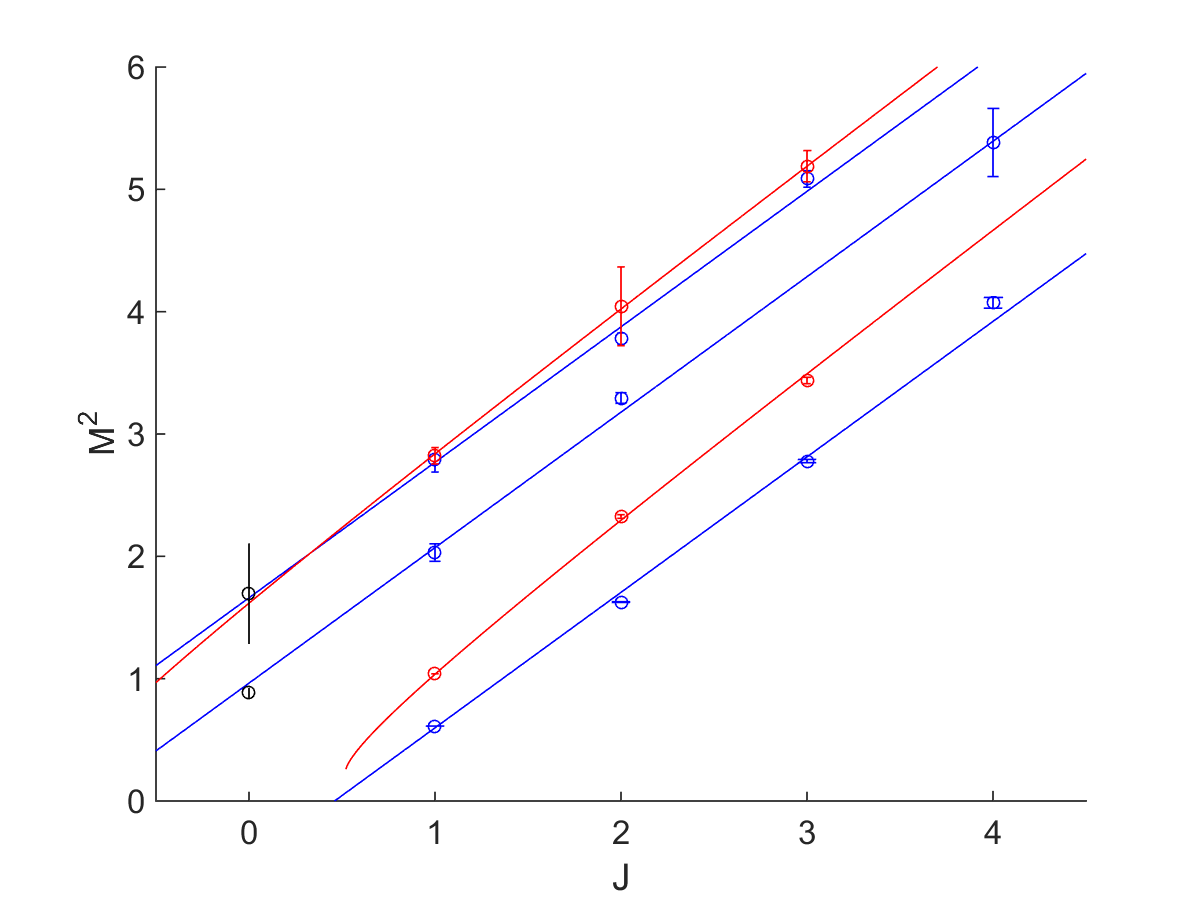}
				\caption{\label{fig:traj_j_mes} The trajectory of the \(\omega\) (blue) and \(\phi\) (red) mesons in the \((J,M^2)\) plane and their daughter trajectories. The fits have the common slope \(\alp = 0.903\) GeV\(^{-2}\), and the \(\ssb\) trajectories are fitted using a mass of \(m_s = 250\) MeV for the \(s\) quark. The states forming the trajectories are as follows: With \(J^{PC} = 1^{--}\), \(\omega(782)\), \(\phi(1020)\), \(\omega(1420)\), \(\omega(1650)\), \(\phi(1680)\). With \(J^{PC} = 2^{++}\), \(f_2(1270)\), \(f_2^\prime(1525)\), \(f_2(1810)\), \(f_2(1950)\), and \(f_2(2010)\). With \(J^{PC} = 3^{--}\), \(\omega_3(1670)\), \(\phi_3(1850)\), \(*\omega_3(2255)\), and \(\omega_3(2285)\). And with \(J^{PC} = 4^{++}\), \(f_4(2050)\) and \(f_4(2300)\). We also plot at \(J^{PC} = 0^{++}\) the \(f_0(980)\) and \(f_0(1370)\) which are found to lie near the trajectories fitted, but were not included themselves in the fits, as they are not theoretically expected to belong to them.}
			\end{figure}

The meson trajectories described above are plotted in figure \ref{fig:traj_j_mes}.

To summarize, we have found several meson trajectories in the \((J,M^2)\) plane of at least three states. As shown in the figure, these trajectories pass quite closely to the states \(f_0(980)\) and \(f_0(1370)\), but as meson trajectories these should begin with a \(J^{PC} = 1^{--}\) state (with orbital angular momentum \(L = 0\) and spin \(S = 1\)). A \(0^{++}\) meson state could only be included as an excited state with \(L = 1\) and \(S = 1\), but we found that for each trajectory we can use an existing \(f_2\) state in that place. The \(f_2\) states classified in this assignment as mesons are \(f_2(1270)\), \(f_2^\prime(1525)\), \(f_2(1810)\), \(f_2(1950)\), and \(f_2(2010)\). These can perhaps be partnered to existing \(f_0\) states as members of triplets of states with \(J = 0, 1, 2\) and \(PC = ++\) split by spin-orbit interactions. We do not know the exact magnitude of the splitting. There are some \(f_0\) states close (within 20--100 MeV) to the \(f_2\) states mentioned above, and the PDG lists some \(f_1\) (\(1^{++}\)) resonances that may be useful, but we do not find any such trio of states with similar properties and masses that could be said to belong to such a spin-orbit triplet. Therefore, we limit our conclusions from these Regge trajectories to the \(f_2\) which we found we could directly place on them.

\subsubsection{\texorpdfstring{Trajectories in the $(n,M^2)$}{(n,M2)} plane} \label{sec:f2_radial}
Sorting the \(f_2\) resonances into trajectories, the situation is somewhat simpler than with the \(f_0\) scalars, as here we have two states that belong on meson trajectories in the \((J,M^2)\) plane, as we found in previous sections. In particular, the \(f_2(1270)\) belongs to the trajectory of the \(\omega\) meson, and the \(f^\prime_2(1525)\) is an \(s\bar{s}\) and sits on the \(\phi\) trajectory. Their decay modes and other properties are also well known and there is no real doubt about their nature.

The linear trajectory beginning with the \(f_2(1270)\) meson includes the states \(f_2(1640)\)and \(f_2(1950)\). We can include one of the further states \(*f_2(2240)\) as the fourth point in the trajectory. We can also use the \(f_J(2220)\) in place of the \(*f_2(2240)\), but it seems an unnatural choice because of the widths of the states involved (the \(f_J(2220)\) is much narrower than the others).

The projected trajectory of the \(f^\prime_2(1525)\), using the same slope as the \(f_2(1270)\) trajectory and adding mass corrections for the \(s\) quark, includes the \(f_2(2010)\) and the \(f_2(2300)\).

This leaves out the states \(f_2(1430)\), \(f_2(1565)\), \(f_2(1810)\), \(f_2(1910)\), \(f_J(2220)\), and \(f_2(2340)\), as well as the five resonances classified as further states.

The next state we look at is \(f_2(1810)\), classified as a light meson in the \((J,M^2)\) fits of the previous section. Its mass is not right for it to belong to the trajectory of the \(f_2(1270)\), so we try to use it as the head of another light meson trajectory. If it belongs to a parallel trajectory to that of the \(f_2(1270)\) then the state that follows it is \(f_2(2150)\). The next state could be \(f_2(2340)\), except that it has been observed to decay to \(\phi\phi\), making it very unlikely to be a light quark meson.

The state \(f_2(1430)\) is intriguing. In part because of the very small width reported by most (but not all) experiments cited in the PDG, and in part because it is located in mass between the two lightest mesons of \(J^{PC} = 2^{++}\), that is between \(f_2(1270)\) (light) and \(f_2^\prime(1525)\) (\(\ssb\)). If we had to assign the \(f_2(1430)\) to a Regge trajectory, then it is best placed preceding the \(f_2(1810)\) and \(f_2(2150)\) in the linear meson trajectory discussed in the last paragraph.

The \(f_J(2200)\), previously known as \(\xi(2230)\), is also a narrow state. It is currently listed by the PDG as having either \(J^{PC} = 2^{++}\) or \(4^{++}\), but some of the experiments cited by the PDG tend towards \(J = 2\). It has been considered a candidate for the tensor glueball \cite{Bai:1996wm,Crede:2008vw}. It can be assigned to a linear meson trajectory, as already discussed, but it is clear already from its narrow width that it is not the best choice, even before addressing other experimental finds regarding it (for example, the fact that it was not observed in \(\gamma\gamma\) scattering \cite{Benslama:2002pa} and the resulting bounds on its decay into photons).

The \(f_2(1565)\) is also left out, but it could be paired with \(f_2(1910)\) to form another linear meson trajectory. To continue we need another state with a mass of around 2200 MeV.

To summarize, we may organize the \(f_2\) resonances by picking first the resonances for the trajectories of the two known mesons,
\[\mathrm{Light:}\qquad 1270, 1640, 1950\]
\[\ssb\mathrm{:}\qquad 1525, 2010, 2300 \]
then find the trajectories starting with the lightest states not yet included. This gives us another meson trajectory using the states
\[\mathrm{Light:}\qquad 1810, 2150\]

\begin{figure}[t!] \centering
	\includegraphics[width=0.76\textwidth]{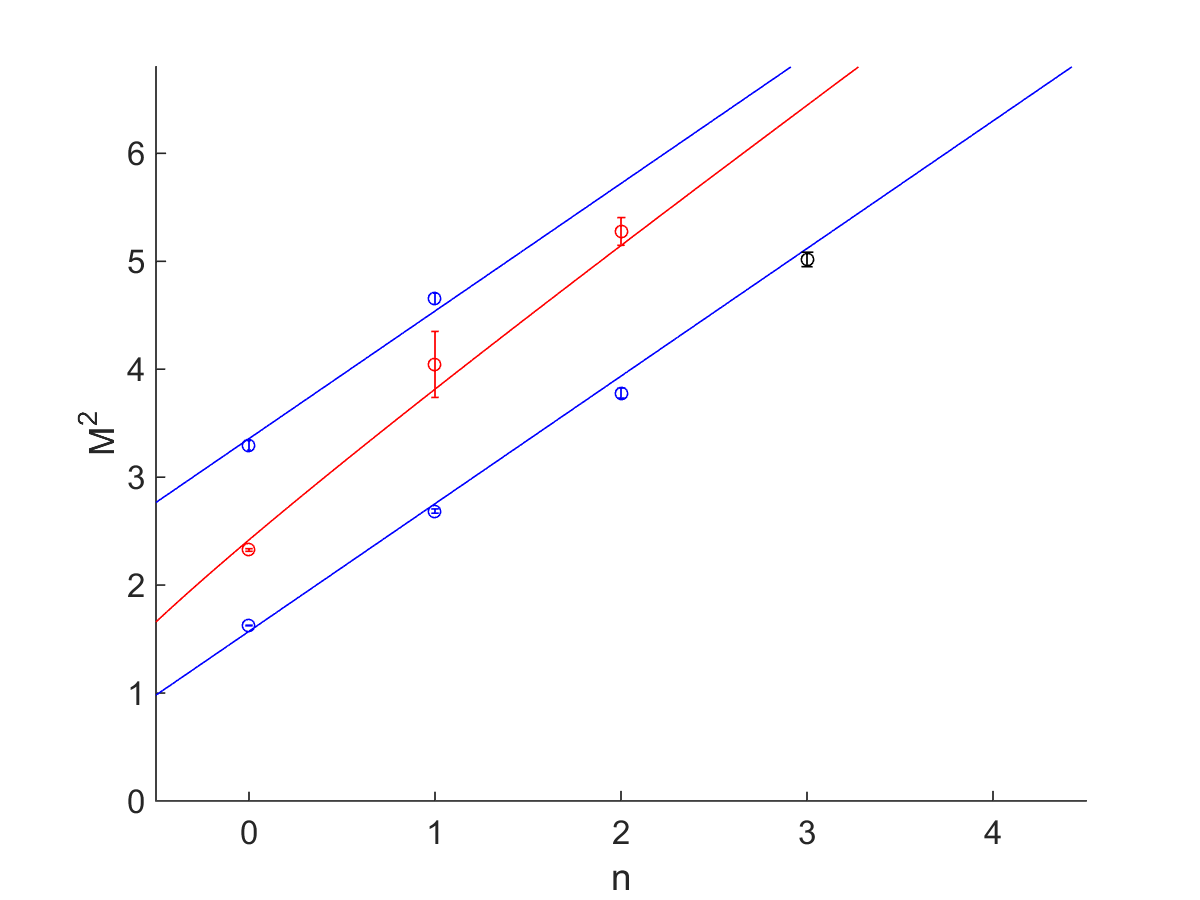}
				\caption{\label{fig:traj_n_f2} Some radial trajectories of the \(f_2\), with blue lines for light mesons and red for \(\ssb\). The fits have the common slope \(\alp = 0.846\) GeV\(^{-2}\), and the \(\ssb\) trajectories are fitted using a mass of \(m_s = 400\) MeV for the \(s\) quark. The states forming the trajectories are as follows: The first light meson trajectory with \(f_2(1270)\), \(f_2(1640)\), and \(f_2(1950)\), and followed by the unconfirmed state \(*f_2(2240)\) which was not used in the fit. The \(\ssb\) trajectory with \(f_2^\prime(1525)\), \(f_2(2010)\), and \(f_2(2300)\). And the second light meson trajectory with \(f_2(1810)\) and \(f_2(2150)\).}
			\end{figure}

The trajectories formed by these eight states are drawn in figure \ref{fig:traj_n_f2}.

\subsubsection{Conclusions from the \texorpdfstring{$f_2$}{f2} fits}
There are some simplifications in assigning the \(f_2\) to radial trajectories compared to assigning the \(f_0\) resonances, as we can look at both orbital and radial trajectories and it is easier to classify some states as mesons. The radial trajectories described in section \ref{sec:f2_radial} are consistent with the orbital trajectories of section \ref{sec:f2_orbital}: states classified as mesons in the latter are also classified as mesons in the former, and with the same quark contents.

In the previous sections we for the most part avoided using the five \(f_2\) states classified in the PDG as ``further states'', although some of them could have played a role in the radial trajectory assignments. Counting confirmed and unconfirmed states alike, the PDG lists a total of 11 states with masses between 1900 and 2340 MeV. Since the different states have been observed in different processes, and hence have different decay modes, it would be useful to clarify experimentally the status of all these states and then reattempt the assignments of the then confirmed states into trajectories. We also note that the fact that there are many resonances with identical quantum numbers near to each other can interfere with the naive mass predictions of the Regge trajectories. In any case, and like for the \(f_0\), further experimental data on resonances between 2.3--3.0 GeV will likely prove useful.

We have not addressed yet the issue of the decay modes of the different states and how consistent they are with the assignments of the previous sections. The \(f_2(1270)\) and \(f_2^\prime(1525)\) are well established as a light quark meson and an \(\ssb\) respectively, and they were the basis from which we built the different trajectories. As for their excited states, the data on their branching ratios cited by the PDG is very partial for higher states. However, we find an interesting case when looking at the trio of states \(f_2(1910)\), \(f_2(1950)\), and \(f_2(2010)\). We have classified \(f_2(1950)\) as a light meson and \(f_2(2010)\) as \(\ssb\), which is what fits best with the Regge trajectories. Another option would be to use \(f_2(1910)\) as a light meson and \(f_2(1950)\) as \(\ssb\), which is still consistent. Then the \(f_2(2010)\), which was observed to decay to \(\phi\phi\) (despite the very small phase space), could perhaps be classified as a \(\phi\phi\) bound state, in an analogous fashion to the \(f_0(980)\).

The most interesting states after that remain the \(f_2(1430)\) and \(f_J(2220)\). While the latter has been considered a candidate for the glueball and has been the object of some research (see papers citing \cite{Bai:1996wm}), the former is rarely addressed, despite its curious placement in the spectrum between the lightest \(2^{++}\) light and \(\ssb\) mesons. It seems a worthwhile experimental question to clarify its status - and its quantum numbers, as the most recent observation \cite{Vladimirsky:2001ek} can not confirm whether it is a \(0^{++}\) or \(2^{++}\) state, a fact which led to at least one suggestion \cite{Vijande:2004he} that the \(f_2(1430)\) could be itself the scalar glueball.

\subsection{Assignments with non-linear trajectories for the glueball} \label{sec:holo_fits}
In this section we check the applicability of a glueball trajectory of the form
\be J = \alp_{gb}E^2-2\alp_{gb}m_0E + a \,, \label{eq:holotraj} \ee
which is the general form we expect from a semi-classical calculation of the corrections to the trajectory in a curved background, and as put forward in section \ref{sec:holo_string}. The novelty here is a term linear in the mass \(E\), which makes the Regge trajectory \(\alpha(t)\) non-linear in \(t = E^2\). The constant \(m_0\) can be either negative or positive, depending on the specific holographic background, and a priori we have to examine both possibilities. It was also noted in section \ref{sec:holo_string} that there may be a correction to the slope, but we assume it is small compared to the uncertainty in the phenomenological value of the Regge slope, and we use
\be \alp_{gb} = \frac{1}{2}\alp \ee
throughout this section. We also substitute \(J \rightarrow J + n\) as usual to apply the formula to radial trajectories.

With the \(m_0\) term we can write
\be \frac{\partial J}{\partial E^2} = \frac{\alp}{2}\left(1-\frac{m_0}{E}\right) \,. \label{eq:eff_holo_slope} \ee
We can look at this as an effective slope, and it is the easiest way to see that when \(m_0\) is negative, the effective slope is higher than that of the linear trajectory, and vice versa.

\subsubsection{Fits using the holographic formula}
Using the simple linear formula we could not, in most cases, find glueball trajectories among the observed \(f_0\) and \(f_2\) states. This is because the first excited state is expected to be too high in mass and outside the range of the states measured in experiment.

Adding an appropriate \(m_0\) term can modify this behavior enough for us to find some pairs of states on what we would then call glueball trajectories, and by appropriate we mean a negative value that will make the effective slope of eq. \ref{eq:eff_holo_slope} higher. The problem is then that we have only pairs of states, with two fitting parameters: \(m_0\) and \(a\) (and \(\alp\) which is fixed by the meson trajectory fits). We form these pairs by picking a state left out from the meson trajectories proposed in sections \ref{sec:f0_fits} and \ref{sec:f2_fits} and assigning it as the excited partner of the appropriate glueball candidate.

There is a solution for \(m_0\) and \(a\) for any pair of states which we can take, and the question then becomes whether there is a reason to prefer some values of the two parameters over others. We list some other values obtained for \(m_0\) and \(a\) in table \ref{tab:holo_fits}.

\begin{table}[t!] \centering
	\begin{tabular}{|c|c|c|c|c|} \hline
	Ground state & Excited state & \(\alp\) [GeV\(^{-2}\)] & $m_0$ [GeV] & \(a\) \\ \hline
	
	\(f_0(980)\) & \(f_0(2200)\) & 0.79 & -0.52 & -0.78 \\
	
	\(f_0(1370)\) & \(f_0(2020)\) & 0.87 & -2.00 & -3.21 \\
	
	\(f_0(1500)\) & \(f_0(2200)\) & 0.87 & -1.51 & -2.97 \\
	
	\(f_2(1430)\) & \(f_J(2220)\) & 0.81 & -1.33 & -2.42 \\ \hline
	
	\end{tabular} \caption{\label{tab:holo_fits} Values obtained for the parameters \(m_0\) and \(a\) for some of the possible pairs of states on glueball trajectories. The states selected as the excited state of the glueball are those not included in the meson trajectories of the assignments of sections \ref{sec:f0_fits} and \ref{sec:f2_fits}, and the slopes are selected based on the results of the meson fits presented in the same sections.}
\end{table}

\subsubsection{Using the holographic formula with a constrained intercept}
\cite{Bigazzi:2004ze} implies that a universal form of the first semi-classical correction of the Regge trajectory of the rotating folded string is
\be J + n = \frac{1}{2}\alp(E-m_0)^2 \,, \ee
up to further (model dependent) modifications of the slope, which in the cases calculated are small. In other words, the intercept obtained then from the semi-classical calculation is
\be a = \frac{1}{2}\alp m_0^2 \,. \ee

The intercept is always positive in this scenario. If we want to include the ground state with \(J = n = 0\) the only way to do it is to take a positive \(m_0\), specifically we should take \(m_0 = M_{gs}\), where \(M_{gs}\) is the mass of the ground state. There is no problem with the resulting expression theoretically, but it is not very useful in analyzing the observed spectrum. The trouble is that when using this expression the energy rises much too fast with \(J\) and we end up very quickly with masses outside the range of the glueball candidates. If we take, for instance, \(f_0(980)\) as the ground state then the first excited state is expected to have a mass of around 2500 MeV, and the heavier candidates naturally predict even heavier masses for the excited states.

Another way to use eq. \ref{eq:holotraj} is to begin the trajectory with a \(J = 2\) state. Then \(m_0\) can be either positive or negative.  We can then proceed as usual: we pick the head of a trajectory and see if there are any matches for its predicted excited states. We can see, for example, that we can again pair \(f_2(1430)\) with \(f_J(2220)\). Constraining \(\alp\) to be \(0.90\) GeV\(^{-2}\), the best fit has \(m_0 = -0.72\) GeV, and the masses calculated are 1390 and 2260 MeV for the experimental values of \(1453\pm4\) and \(2231\pm4\) MeV.

\subsection{Glueball Regge trajectories in lattice QCD} \label{sec:lattice}
The glueball spectrum has been studied extensively in lattice QCD. Some works have compared results with different stringy models, e.g. \cite{Athenodorou:2010cs,Bochicchio:2013aha,Bochicchio:2013sra,Caselle:2015tza}. However, the question whether or not the glueballs form linear Regge trajectories is not often addressed, due to the difficulty involved in computing highly excited states. When linear Regge trajectories are discussed, it is often when trying to identify the glueball with the pomeron and searching for states along the given pomeron trajectory,
\be \alpha(t) = \alp_p t + 1 + \epsilon \, \ee
 where the slope and the intercept are known from experiment to be \(\alp_p = 0.25\) GeV\(^{-2}\) and \(1 + \epsilon \approx 1.08\) \cite{Donnachie:1984xq}.

The most extensive study of glueball Regge trajectories is that of Meyer and Teper \cite{Meyer:2004jc,Meyer:2004gx}, where a relatively large number of higher mass states is computed, including both high spin states and some highly excited states at low spin.

We quote in table \ref{tab:lat_masses} some lattice results for glueball masses from different calculations. The results are for \(SU(3)\) and \(D = 4\), and more results are collected in \cite{Gregory:2012hu}. Most of these give only the masses of the lowest glueball states for different quantum numbers. These are low spin states with different combinations of parity and charge parity. While a spectrum is obtained, most states are isolated, in the sense that they cannot be grouped with other states to form Regge trajectories.

In the table \ref{tab:lat_masses} we list the lattice results for the \(0^{++}\) ground state, the lowest \(2^{++}\) state, and the first excited \(0^{++}\) glueball, as well as for the \(0^{-+}\) and \(2^{-+}\). We may straight lines between the first spin-0 state and its excited partner to calculate the slope.

One thing we see at this first glance at the spectrum is that the spin-2 state is, in most studies, lower than we would expect it based on the Regge slope assumption.\footnote{The fact that the tensor glueball is close to the scalar seems to have been long known in lattice QCD, see e.g. \cite{Albanese:1987ds}.} The second spin-0 state, on the other hand, is about where we want it to be, assuming a closed string model, where the slope is half that of meson trajectories, and the first excited state has the excitation number \(n = 2\) (for one left moving and one right moving mode excited). In the next section we do some fits to some trajectories with more than two states, based on the results in \cite{Meyer:2004gx}.

\begin{table}[t!] \centering
	\begin{tabular}{|c|c|c|c|c|c|c|} \hline
	& Meyer \cite{Meyer:2004gx}  & M\&P \cite{Morningstar:1999rf} & Chen \cite{Chen:2005mg} & Bali \cite{Bali:1993fb} & Gregory \cite{Gregory:2012hu} \\ \hline\hline

	\(0^{++}\) & 1475\plm30\plm65			& 1730\plm50\plm80 & 1710\plm50\plm80 & 1550\plm50\plm80 & 1795\plm60 \\ \hline

	\(2^{++}\) & 2150\plm30\plm100		& 2400\plm25\plm120 & 2390\plm30\plm120 & 2270\plm100\plm110 & 2620\plm50\\ \hline

	\(0^{++}\) & 2755\plm30\plm120		& 2670\plm180\plm130 & - & - & 3760\plm240 \\ \hline\hline

	\(0^{-+}\) & 2250\plm60\plm100			& 2590\plm40\plm130 & 2560\plm35\plm120 & 2330\plm260\plm120 & - \\ \hline

	\(2^{-+}\) & 2780\plm50\plm130		& 3100\plm30\plm150 & 3040\plm40\plm150 & 3010\plm130\plm150 & 3460\plm320\\ \hline

	\(0^{-+}\) & 3370\plm150\plm150		& 3640\plm60\plm180 & - & - & 4490\plm590 \\ \hline\hline

	\(\alp\!_{++}\) (in $J$) & 0.82\plm0.17 & 0.72\plm0.18 & 0.72\plm0.17 & 0.73\plm0.19 & 0.55\plm0.05 \\ \hline

	\(\alp\!_{++}\) (in $n$) & 0.37\plm0.05 & 0.48\plm0.14 & - & - & 0.18\plm0.03 \\ \hline\hline

	\(\alp\!_{-+}\) (in $J$) & 0.75\plm0.26 & 0.69\plm0.28 & 0.74\plm0.32 & 0.55\plm0.27 & - \\ \hline

	\(\alp\!_{-+}\) (in $n$) & 0.32\plm0.08 & 0.31\plm0.07 & - & - & - \\ \hline
	\end{tabular}
	\caption{\label{tab:lat_masses} Lattice predictions from different studies for glueball masses [MeV] and resulting Regge slopes [GeV\(^{-2}\)], in the \((J,M^2)\) plane or in the \((n,M^2)\) plane. The slope is calculated assuming the first excited state has \(n = 2\).}
	\end{table}

\subsubsection{Regge trajectory fits to results from the lattice}
Results in lattice computations are for the dimensionless ratio between the mass of a state and the square root of the string tension: \(M/\sqrt{T}\). To get the masses \(M\) in MeV one has to fix the scale by setting the value of \(T\). This introduces an additional uncertainty in the obtained values. In table \ref{tab:lat_masses} we listed the masses in MeV and calculated the dimensionful slope, but for the purpose of identifying Regge trajectories we can work directly with dimensionless quantities, avoiding this extra error. Thus, for the following, our fitting model will be
\be \frac{M^2}{T} = \frac{2\pi}{q} (N + a) \ee
In this notation the ratio \(q\), which is the primary fitting parameter (in addition to the intercept \(a\)), is expected to be 1 for open strings and \(1/2\) for closed strings. It is referred to below as the ``relative slope''. \(N\) will be either the spin \(J\) or the radial excitation number \(n\).

\begin{figure}[tp!] \centering
	\includegraphics[width=0.49\textwidth]{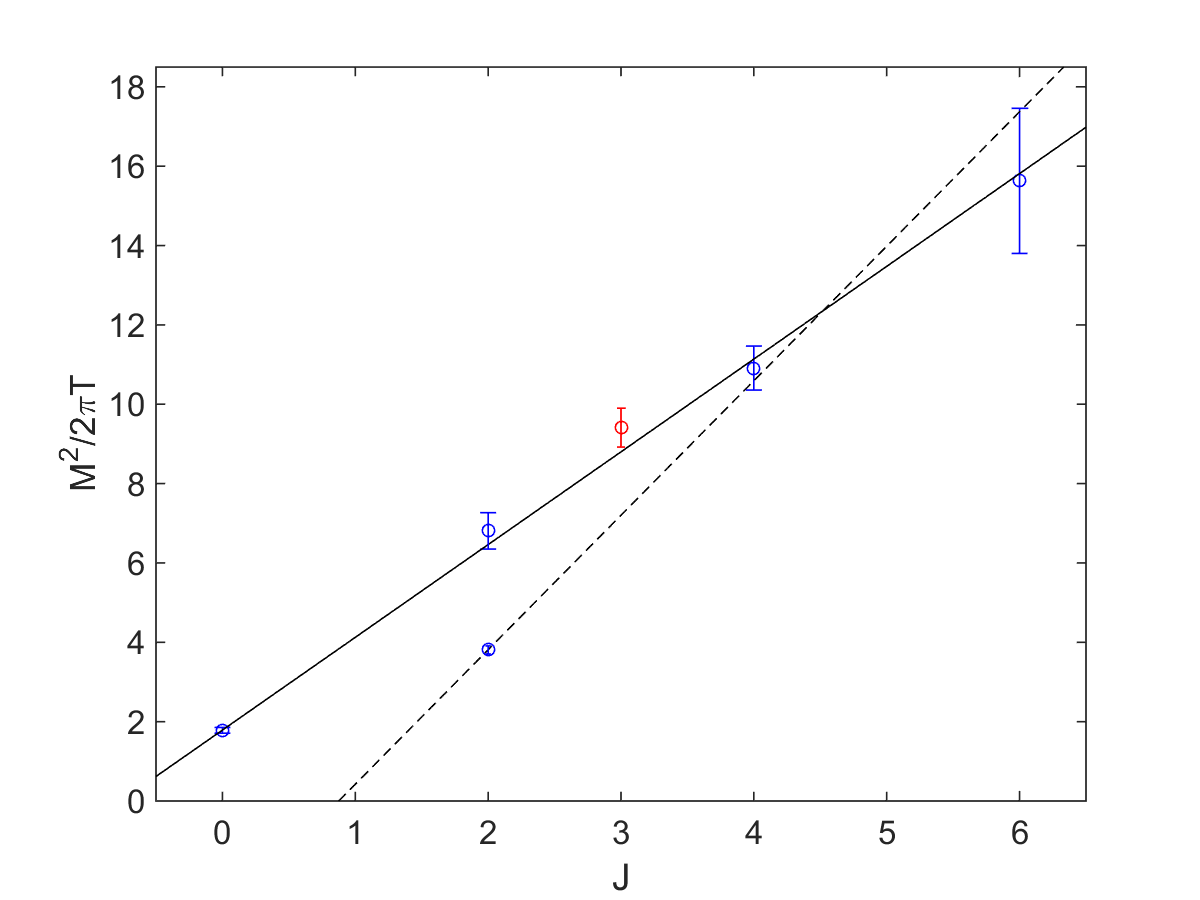}
	\includegraphics[width=0.49\textwidth]{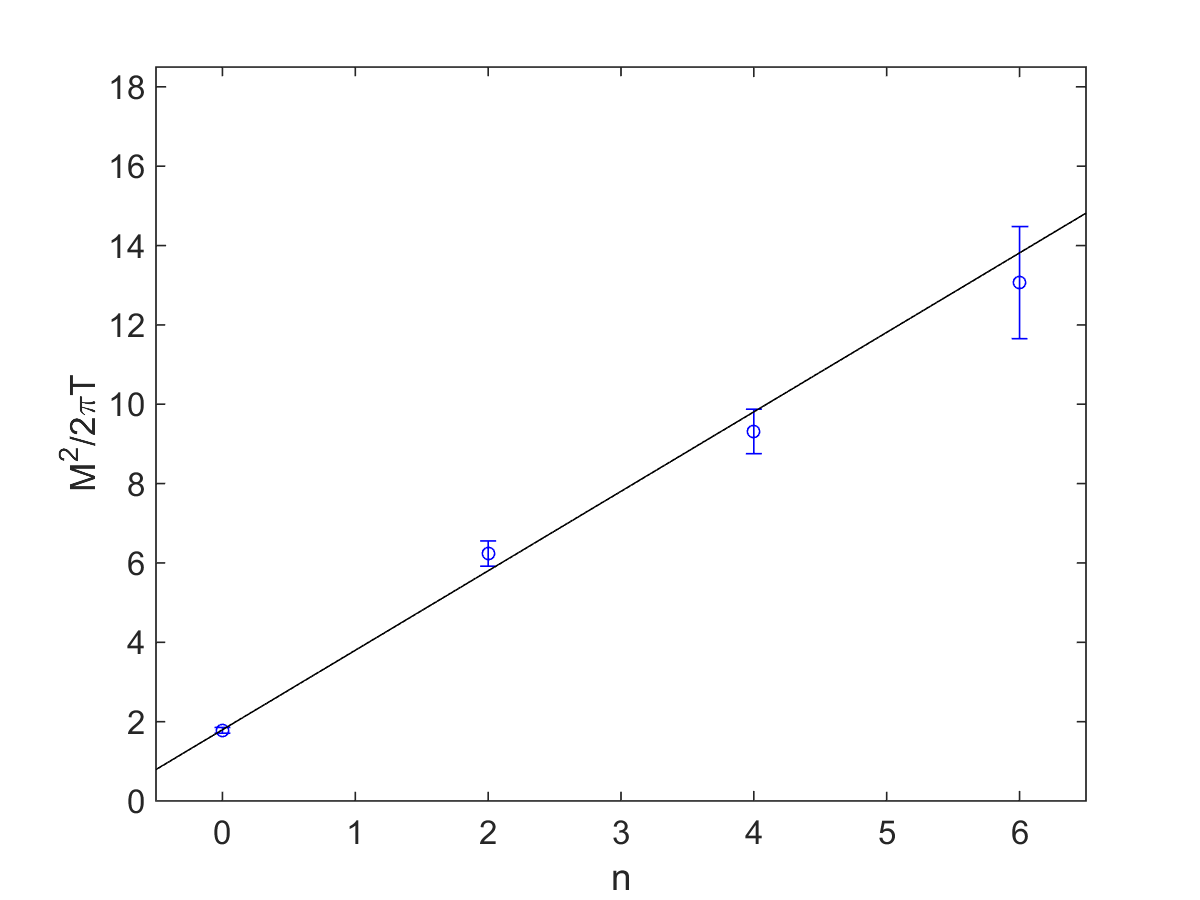}
	\caption{\label{fig:lat_Meyer} The trajectories of the \(PC = ++\) glueball states found in lattice calculations in \cite{Meyer:2004gx}. \textbf{Left:} Trajectories in the \((J,M^2)\) plane. The full line is the fit to a proposed trajectory using four states with \(J = 0, 2, 4, 6\), where the relative slope is \(0.43\) and the lightest tensor is excluded (\(\chi^2 = 0.37\)). The dotted line is the leading trajectory proposed in the analysis in \cite{Meyer:2004gx}, with a pomeron-like slope. It includes the \(J = 2, 4,\) and \(6\) states. (\(\chi^2 = 1.24\)). In this second option the scalar is excluded. Also plotted is the \(3^{++}\) state, which was not used in the fit. \textbf{Right:} trajectory of four states with \(J^{PC} = 0^{++}\). The relative slope is exactly \(0.50\) (\(\chi^2 = 1.48\)).}
\end{figure}

\paragraph{Trajectories in the \texorpdfstring{$(J,M^2)$}{(J,M2)} plane:} As mentioned above, \cite{Meyer:2004gx} has the most high spin states. The analysis there observes that the first \(2^{++}\)and \(4^{++}\) states can be connected by a line with the relative slope
\be q = 0.28\pm0.02, \ee
which, when taking a typical value of the string tension \(\sqrt{T} = 430\) MeV (\(\alp = 0.84\) GeV\(^{-2}\)), gives a slope virtually identical to that expected for the pomeron, \(0.25\) GeV\(^{-2}\). This trajectory can be continued with the calculated \(6^{++}\) state. A fit to the three state trajectory gives the result
\be q = 0.29\pm0.15.\ee
This trajectory leaves out the \(0^{++}\) ground state. In \cite{Meyer:2004gx} the lowest \(0^{++}\) is paired with the second, excited, \(2^{++}\) state, giving a trajectory with
\be q = 0.40\pm0.04.\ee
A possibility not explored in \cite{Meyer:2004gx} is that of continuing this trajectory, of the first \(0^{++}\) and the excited \(2^{++}\), and with the \(4^{++}\) and \(6^{++}\) states following. Then we have the result
\be q = 0.43\pm0.03\ee
This second option not only includes more points, it is also a better fit in terms of \(\chi^2\) per degrees of freedom (0.37 instead of 1.24). This is a nice result from the closed string perspective, but the lowest \(2^{++}\) state is then left out. There is also a \(J = 3\) state in the \(PC = ++\) sector that lies very close to the trajectory of the \(0^{++}\) ground state. In our model it is not expected to belong to the trajectory, so that state is also left out of the fit. The trajectories of the \(PC = ++\) states are in the left side of figure \ref{fig:lat_Meyer}.

\paragraph{Trajectories in the \texorpdfstring{$(n,M^2)$}{(n,M2)} plane:}
In trajectories in the \((n,M^2)\) plane we assume \(n\) takes only even values, i.e. \(n = 0,2,4,\ldots\), as it does for the closed string. The results when taking \(n = 0,1,2,\ldots\) will be half those listed.

In this section, we again have to rely mostly on \cite{Meyer:2004gx}, as it offers calculations of several excited states with the same \(J^{PC}\). Most notably we see there four states listed with \(J^{PC} = 0^{++}\). We observe that those points are well fitted by a trajectory with the slope
\be q = 0.50\plm0.07, \ee
where \(\chi^2 = 1.48\) for the fit. It is interesting to compare this with the trajectory that can be drawn from the \(0^{++}\) ground state in the \((J,M^2)\) plane. The \((n,M^2)\) trajectory with \(n = 0,2,4,6\) is very similar to the trajectory beginning with the same state and continuing to \(J = 2, 4,\) and \(6\). This is what we see also for mesons and baryons in experiment: two analogous trajectories with similar slopes in the different planes.

Other than the trajectory of the four \(0^{++}\) states (plotted in figure \ref{fig:lat_Meyer}), we list the slopes calculated for pairs of states who share other quantum numbers. This is in table \ref{tab:lat_n}.
\begin{table} \centering
	\begin{tabular}{|c|c|c|c|c|c|} \hline
		\(J^{PC}\) & \(0^{++}\) & \(2^{++}\) & \(4^{++}\) & \(0^{-+}\) & \(2^{-+}\) \\ \hline\hline
		Meyer \cite{Meyer:2004gx} &
				0.50\plm0.07 & 0.67\plm0.10 & 0.30\plm0.06 & 0.39\plm0.07 & 0.56\plm0.13 \\ \hline

		M\&P \cite{Morningstar:1999rf} &
				0.51\plm0.12 & -						& -						 & 0.32\plm0.02 & 0.38\plm0.03 \\ \hline
	\end{tabular}
	\caption{\label{tab:lat_n} Relative slopes \(q\) of trajectories in the \((n,M^2)\) plane. The first result (Meyer/\(0^{++}\)) is that of a fit to the four point trajectory drawn in \ref{fig:lat_Meyer}. The other results are obtained when calculating the slopes between pairs of states, where the lowest state is assumed to have \(n = 0\), and the first excited state is taken to have \(n = 2\).}
\end{table}

\subsubsection{\texorpdfstring{$SU(N)$ vs. $SU(3)$}{SU(N) vs. SU(3)} and the quenched approximation}
Most of the studies of glueballs on the lattice utilize the ``quenched'' approximation, which in this case amounts to calculating the spectrum of the pure SU(3) Yang-Mills gauge theory without matter. The degree to which the quenched results are modified when fermions are added to the theory is still unknown. However, if our purpose is to see whether or not glueballs form Regge trajectories, the spectrum of the pure gluon theory should be as useful as that of real QCD.

There have also been some calculations of the ``glueball'' spectrum of \(SU(N)\) Yang-Mills for other values of \(N\) \cite{Lucini:2014paa}. These results taken from \cite{Lucini:2004my}, are fitted in \cite{Lucini:2014paa} to the formulae (the numbers in brackets are the errors in the last significant digits):
\be \frac{M_{0^{++}}}{\sqrt{T}} = 3.28(8)+\frac{2.1(1.1)}{N^2} \,,\ee
\[ \frac{M_{2^{++}}}{\sqrt{T}} = 4.78(14)+\frac{0.3(1.7)}{N^2} \,,\]
\[ \frac{M_{0^{++*}}}{\sqrt{T}} = 5.93(17)-\frac{2.7(2.0)}{N^2} \,.\]
Using these values, we get for \(SU(3)\), the relative slopes (the prefactor of 2 in these formulae is \(J\) or \(n\)):
\be 2\frac{2\pi T}{M^2_{2^{++}}-M^2_{0^{++}}} = 1.16\plm0.27, \qquad
 2\frac{2\pi T}{M^2_{0^{++*}}-M^2_{0^{++}}} = 0.65\plm0.11 \,, \ee
while for the \(N\rightarrow\infty\) limit,
\be 2\frac{2\pi T}{M^2_{2^{++}}-M^2_{0^{++}}} = 1.04\plm0.13, \qquad
 2\frac{2\pi T}{M^2_{0^{++*}}-M^2_{0^{++}}} = 0.52\pm0.05 \,.\ee

While this is too little data to be significant, we observe that the value approaches 1 as \(N\) grows for the excitation in \(J\), and it approaches \(\frac12\) (the closed string value) in \(n\). However, as was already seen from the results of \cite{Meyer:2004gx}, the first \(2^{++}\) does not seem to lie on the trajectory of the \(0^{++}\) ground state, and these results seem to confirm this further. The radial trajectory on the other hand is again perfectly consistent with the closed string picture, and more so when going to the limit of this large \(N\) computation.

\section{Summary} \label{sec:summary}

For many years the identification of glueballs, a basic prediction of QCD, in the experimental spectrum of flavorless isoscalar hadrons has been an open question.  Moreover, the common lore is that there is no way to disentangle glueballs from flavorless mesons since there is no quantum number that distinguish between them. 

Here in this paper we have attempted to identify glueballs by turning to a well known feature of the hadron spectrum, its Regge trajectories. Stating it differently we use a stringy picture of rotating folded closed strings to describe the glueball in a similar way to the description of mesons and baryons in terms of open string with massive endpoints of \cite{Sonnenschein:2014jwa} and \cite{Sonnenschein:2014bia}.

The great disadvantage in using trajectories is that they are a property not of single states, but of a spectrum of states. Thus, for positive identification, we need to have in our spectrum, to begin with, several glueballs which we would then assign to a trajectory. The fact that the ratio between the open and closed string slopes is exactly half adds some ambiguity to the \((n,M^2)\) trajectories where the value of \(n\) cannot be determined by experiment: two states whose mass difference is, for instance, \(\Delta M^2 = 4/\alp\) can be either open strings with \(\Delta n = 4\) between them, or closed strings with \(\Delta n = 2\). The difference between the open and closed string trajectories would be in the number of states between those two: there would be more open strings for \(\Delta n = 1\), 2, and 3. Thus we have to rely on experiment to observe all the relevant states in the given mass range, so that the absence of a state from a Regge trajectory could reasonably be used as evidence.

Due to this situation it is clearly advisable to use additional predictions pertaining to the properties of single states to identify them as open or closed string hadrons. We have presented, qualitatively, the decay mechanism of the closed string to two open strings, which would be the decay of a glueball into two mesons. We included one prediction of the branching ratios of glueballs when decaying into light mesons, kaons, or \(\phi\) (\(s\bar{s}\)) mesons. If there were measurements of a state which has those three decay modes with the hierarchy we predict between them, we could have declared it a glueball, based on our model of holographic strings. One has to look more closely to find more ways in which open and closed strings vary.

There are obviously additional tasks and questions to further explore the closed string picture of glueballs. Here we list some of them:
\begin{itemize}
\item
As was emphasized in this note the most urgent issue is to gain additional data about flavorless hadrons. This calls for a further investigation of experiments that yield this kind of resonances and for proposing future experiments of potential glueball production, in particular in the range above 2.4 GeV. This can follow the predictions of the masses and width of the resonances as were listed in appendix \ref{sec:predictions}.
\item
Related to the exploration of experimental data is the investigation of efficient mechanisms of creating glueballs. This issue was not addressed in this paper. Among possible glueball formation one finds radiative $J/\psi$ decays, pomeron pomeron collisions in hadron-hadron central production and in $p$-$\bar p$ annihilation. Naturally, we would like to understand possible glueball formation in LHC experiments. It is known that we can find in the latter processes of gluon-gluon scattering and hence it may serve as a device for glueball creation. 
\item
As was mentioned in section \ref{sec:non_critical_string}, the quantization of folded closed strings in D non-critical dimensions has not yet been deciphered. In \cite{Hellerman:2013kba} the expression derived for the intercept is singular in the case where is only one rotation plane - as it naturally is in $D=4$. We mentioned a potential avenue to resolve this issue by introducing massive particles on the folds, quantize the system as that of a string with massive endpoints \cite{ASY}, and then take the limit of zero mass. 
\item
We have mentioned that the rotating closed strings are in fact rotating folded closed strings. However, we did not make any attempt in this note to explore the role of the folds. In fact it seems that very few research has been devoted to the understanding of folded strings \cite{Ganor:1994rm}. It would be interesting to use the rotating closed string as a venue to the more general exploration of strings with folds which may be related to certain systems in nature.
\item
A mystery related to the closed string description of glueballs is the relation between the pomeron and the glueball. Supposedly  both the glueball and the pomeron are described by a closed string. As we have emphasized in this note the slope of the closed string is half that of the open string and hence we advocated the search of trajectories with that slope. However, it was found from fitting the differential cross section of $p$-$p$ collisions that the slope of the pomeron is $\alp_{pomeron}\approx 0.25$ GeV$^{-2}$. That is, a slope which is closer to a quarter of that of the meson open string rather than half. Thus the stringy structure of the pomeron and its exact relation to the glueball is still an open question.
\item
The closed string description of the glueball faces a very obvious question. In QCD one can form a glueball as a bound state of two, three, or in fact any number of gluons. The stringy picture seems to describe the composite of two gluons, and it is not clear how to realize those glueballs constructed in QCD from more than two gluons.
\end{itemize} 

\acknowledgments{
We would like to thank Ofer Aharony and Abner Soffer for their comments on the manuscript and for insightful conversations, and Shmuel Nussinov and Shimon Yankielowicz for useful discussions.  This work was supported in part by a centre of excellence supported by the Israel Science Foundation (grant  number 1989/14), and by the US-Israel bi-national fund (BSF) grant number 2012383 and the Germany Israel bi-national fund GIF grant number I-244-303.7-2013. 
}

\appendix

\section{Predictions} \label{sec:predictions}

\subsection{Predictions for glueballs}
In this section we list the masses obtained when using linear trajectories for glueballs, with an appropriate slope \(\alp_{gb} = \frac12\alp_{meson}\), and based on assigning one of the candidates as the ground state for the trajectory.

The slope \(\alp_{gb}\) is taken to be 0.40--0.45 GeV\(^{-2}\). This is based on the values for the light meson slopes of the fits in \cite{Sonnenschein:2014jwa}, as well as on the results of the fits in section \ref{sec:phenomenology} of this paper. In \cite{Sonnenschein:2014jwa} we see that in general the slopes of the \((J,M^2)\) trajectories tend to be higher than those of the radial \((n,M^2)\) trajectories. The typical values are 0.90 GeV\(^{-2}\) for the former, and closer to 0.80 GeV\(^{-2}\) for the latter. In the radial fits done in this paper we get slopes between 0.79 and 0.88 GeV\(^{-2}\), depending on assignments. The predictions are not based specifically on these assignments, but we maintain that the range for the slope mentioned above \(\alp_{gb} = \) 0.40--0.45 GeV\(^{-2}\) is valid for both the \((J,M^2)\) and \((n,M^2)\) trajectories of the glueball.

We also include a prediction for the widths of the excited states. When calculating the width, we assume the simple relation \(\Gamma/M^2 = Const.\) as put forward in section \ref{sec:decays}. Therefore the width of a state with mass \(M\) is calculated using
\be \Gamma = M^2\frac{\Gamma_0}{M_0^2}\,, \ee
with \(M_0\) and \(\Gamma_0\) the experimentally measured mass and width of the ground state.

The error bars take into account both the experimental uncertainty in \(M_0\) and \(G_0\) and the uncertainty in the parameter \(\alp_{gb}\).

We present the results in tables \ref{tab:pred980}--\ref{tab:pred1710}.

\begin{table}[htp!] \centering
	\begin{tabular}{|c|c|c|} \hline
\(n\) or \(J\) 	&	 Mass 	&	 Width \\ \hline
0 	&	990\plm20 	&	 70\plm30	\\ \hline
2	&	2385\plm70 	&	 405\plm175	\\ \hline
4	&	3225\plm95 	&	 740\plm325	\\ \hline
6	&	3885\plm115 	&	 1080\plm470	\\ \hline
8	&	4450\plm130 	&	 1415\plm615	\\ \hline
\end{tabular} \caption{Predictions using linear trajectories with glueball slope \(\alp_{gb} = \) 0.40--0.45 GeV\(^{-2}\) using \(f_0(980)\) as the ground state. All states have \(PC = ++\), and one of \(n\) or \(J\) should be taken as zero along a trajectory, as the other changes. \label{tab:pred980}}
\end{table}

\begin{table}[htp!] \centering
	\begin{tabular}{|c|c|c|} \hline
\(n\) or \(J\) 	&	 Mass 	&	 Width \\ \hline
0 	&	1350\plm150 	&	 350\plm150	\\ \hline
2	&	2555\plm110 	&	 1255\plm615	\\ \hline
4	&	3350\plm115 	&	 2155\plm1050	\\ \hline
6	&	3995\plm130 	&	 3060\plm1490	\\ \hline
8	&	4545\plm140 	&	 3965\plm1930	\\ \hline
\end{tabular} \caption{Predictions using linear trajectories with glueball slope \(\alp_{gb} = \) 0.40--0.45 GeV\(^{-2}\) using \(f_0(1370)\) as the ground state. All states have \(PC = ++\), and one of \(n\) or \(J\) should be taken as zero along a trajectory, as the other changes. \label{tab:pred1370}}
\end{table}

\begin{table}[htp!] \centering
	\begin{tabular}{|c|c|c|} \hline
\(n\) or \(J\) 	&	 Mass 	&	 Width \\ \hline
0 	&	1505\plm6 	&	 109\plm7	\\ \hline
2	&	2640\plm80 	&	 335\plm30	\\ \hline
4	&	3415\plm100 	&	 560\plm50	\\ \hline
6	&	4050\plm120 	&	 790\plm70	\\ \hline
8	&	4590\plm135 	&	 1015\plm90	\\ \hline
\end{tabular} \caption{Predictions using linear trajectories with glueball slope \(\alp_{gb} = \) 0.40--0.45 GeV\(^{-2}\) using \(f_0(1500)\) as the ground state. All states have \(PC = ++\), and one of \(n\) or \(J\) should be taken as zero along a trajectory, as the other changes. \label{tab:pred1500}}
\end{table}

\begin{table}[htp!] \centering
	\begin{tabular}{|c|c|c|} \hline
\(n\) or \(J\) 	&	 Mass 	&	 Width \\ \hline
0 	&	1720\plm6 	&	 135\plm8	\\ \hline
2	&	2770\plm85 	&	 350\plm30	\\ \hline
4	&	3515\plm105 	&	 565\plm50	\\ \hline
6	&	4130\plm125 	&	 780\plm65	\\ \hline
8	&	4665\plm140 	&	 995\plm85	\\ \hline
\end{tabular} \caption{Predictions using linear trajectories with glueball slope \(\alp_{gb} = \) 0.40--0.45 GeV\(^{-2}\) using \(f_0(1710)\) as the ground state. All states have \(PC = ++\), and one of \(n\) or \(J\) should be taken as zero along a trajectory, as the other changes. \label{tab:pred1710}}
\end{table}

\clearpage

\bibliographystyle{JHEP}
\bibliography{Glueballs}

\providecommand{\href}[2]{#2}\begingroup\raggedright\begin{thebibliography}{10}

\bibitem{Sonnenschein:2014jwa}
J.~Sonnenschein and D.~Weissman, {\it {Rotating strings confronting PDG
  mesons}},  {\em JHEP} {\bf 1408} (2014) 013,
  [\href{http://xxx.lanl.gov/abs/1402.5603}{{\tt arXiv:1402.5603}}].

\bibitem{Sonnenschein:2014bia}
J.~Sonnenschein and D.~Weissman, {\it {A rotating string model versus baryon
  spectra}},  {\em JHEP} {\bf 1502} (2015) 147,
  [\href{http://xxx.lanl.gov/abs/1408.0763}{{\tt arXiv:1408.0763}}].

\bibitem{Csaki:1998qr}
C.~Csaki, H.~Ooguri, Y.~Oz, and J.~Terning, {\it {Glueball mass spectrum from
  supergravity}},  {\em JHEP} {\bf 9901} (1999) 017,
  [\href{http://xxx.lanl.gov/abs/hep-th/9806021}{{\tt hep-th/9806021}}].

\bibitem{Brower:2000rp}
R.~C. Brower, S.~D. Mathur, and C.-I. Tan, {\it {Glueball spectrum for QCD from
  AdS supergravity duality}},  {\em Nucl. Phys.} {\bf B587} (2000) 249--276,
  [\href{http://xxx.lanl.gov/abs/hep-th/0003115}{{\tt hep-th/0003115}}].

\bibitem{Elander:2013jqa}
D.~Elander, A.~F. Faedo, C.~Hoyos, D.~Mateos, and M.~Piai, {\it {Multiscale
  confining dynamics from holographic RG flows}},  {\em JHEP} {\bf 05} (2014)
  003, [\href{http://xxx.lanl.gov/abs/1312.7160}{{\tt arXiv:1312.7160}}].

\bibitem{Bhanot:1980fx}
G.~Bhanot and C.~Rebbi, {\it {SU(2) String Tension, Glueball Mass and
  Interquark Potential by Monte Carlo Computations}},  {\em Nucl.Phys.} {\bf
  B180} (1981) 469.

\bibitem{Niemi:2003hb}
A.~J. Niemi, {\it {Are glueballs knotted closed strings?}},
  \href{http://xxx.lanl.gov/abs/hep-th/0312133}{{\tt hep-th/0312133}}.

\bibitem{Sharov:2007ag}
G.~Sharov, {\it {Closed String with Masses in Models of Baryons and
  Glueballs}},  \href{http://xxx.lanl.gov/abs/0712.4052}{{\tt
  arXiv:0712.4052}}.

\bibitem{Solovev:2000nb}
L.~Solovev, {\it {Glueballs in the string quark model}},  {\em
  Theor.Math.Phys.} {\bf 126} (2001) 203--211,
  [\href{http://xxx.lanl.gov/abs/hep-ph/0006010}{{\tt hep-ph/0006010}}].

\bibitem{Talalov:2001cp}
S.~Talalov, {\it {The Glueball Regge trajectory from the string inspired
  theory}},  \href{http://xxx.lanl.gov/abs/hep-ph/0101028}{{\tt
  hep-ph/0101028}}.

\bibitem{LlanesEstrada:2000jw}
F.~J. Llanes-Estrada, S.~R. Cotanch, P.~J. de~A.~Bicudo, J.~E.~F. Ribeiro, and
  A.~P. Szczepaniak, {\it {QCD glueball Regge trajectories and the Pomeron}},
  {\em Nucl.Phys.} {\bf A710} (2002) 45--54,
  [\href{http://xxx.lanl.gov/abs/hep-ph/0008212}{{\tt hep-ph/0008212}}].

\bibitem{Szczepaniak:2003mr}
A.~P. Szczepaniak and E.~S. Swanson, {\it {The Low lying glueball spectrum}},
  {\em Phys.Lett.} {\bf B577} (2003) 61--66,
  [\href{http://xxx.lanl.gov/abs/hep-ph/0308268}{{\tt hep-ph/0308268}}].

\bibitem{Pons:2004dk}
J.~Pons, J.~Russo, and P.~Talavera, {\it {Semiclassical string spectrum in a
  string model dual to large N QCD}},  {\em Nucl.Phys.} {\bf B700} (2004)
  71--88, [\href{http://xxx.lanl.gov/abs/hep-th/0406266}{{\tt
  hep-th/0406266}}].

\bibitem{Abreu:2005uw}
E.~Abreu and P.~Bicudo, {\it {Glueball and hybrid mass and decay with string
  tension below Casimir scaling}},  {\em J.Phys.} {\bf G34} (2007) 195207,
  [\href{http://xxx.lanl.gov/abs/hep-ph/0508281}{{\tt hep-ph/0508281}}].

\bibitem{Brau:2004xw}
F.~Brau and C.~Semay, {\it {Semirelativistic potential model for glueball
  states}},  {\em Phys.Rev.} {\bf D70} (2004) 014017,
  [\href{http://xxx.lanl.gov/abs/hep-ph/0412173}{{\tt hep-ph/0412173}}].

\bibitem{Mathieu:2005wc}
V.~Mathieu, C.~Semay, and F.~Brau, {\it {Casimir scaling, glueballs and hybrid
  gluelumps}},  {\em Eur.Phys.J.} {\bf A27} (2006) 225--230,
  [\href{http://xxx.lanl.gov/abs/hep-ph/0511210}{{\tt hep-ph/0511210}}].

\bibitem{Simonov:2006re}
Y.~Simonov, {\it {Glueballs, gluerings and gluestars in the d=2+1 SU(N) gauge
  theory}},  {\em Phys.Atom.Nucl.} {\bf 70} (2007) 44--52,
  [\href{http://xxx.lanl.gov/abs/hep-ph/0603148}{{\tt hep-ph/0603148}}].

\bibitem{Mathieu:2006bp}
V.~Mathieu, C.~Semay, and B.~Silvestre-Brac, {\it {Semirelativistic potential
  model for low-lying three-gluon glueballs}},  {\em Phys.Rev.} {\bf D74}
  (2006) 054002, [\href{http://xxx.lanl.gov/abs/hep-ph/0605205}{{\tt
  hep-ph/0605205}}].

\bibitem{BoschiFilho:2002vd}
H.~Boschi-Filho and N.~R. Braga, {\it {Gauge / string duality and scalar
  glueball mass ratios}},  {\em JHEP} {\bf 0305} (2003) 009,
  [\href{http://xxx.lanl.gov/abs/hep-th/0212207}{{\tt hep-th/0212207}}].

\bibitem{Peeters:2005fq}
K.~Peeters, J.~Sonnenschein, and M.~Zamaklar, {\it {Holographic decays of
  large-spin mesons}},  {\em JHEP} {\bf 0602} (2006) 009,
  [\href{http://xxx.lanl.gov/abs/hep-th/0511044}{{\tt hep-th/0511044}}].

\bibitem{PandoZayas:2003yb}
L.~A. Pando~Zayas, J.~Sonnenschein, and D.~Vaman, {\it {Regge trajectories
  revisited in the gauge / string correspondence}},  {\em Nucl.Phys.} {\bf
  B682} (2004) 3--44, [\href{http://xxx.lanl.gov/abs/hep-th/0311190}{{\tt
  hep-th/0311190}}].

\bibitem{Polchinski:Vol1}
J.~Polchinski, {\it {String theory. Vol. 1: An introduction to the bosonic
  string}}, .

\bibitem{Hellerman:2013kba}
S.~Hellerman and I.~Swanson, {\it {String Theory of the Regge Intercept}},
  {\em Phys.Rev.Lett.} {\bf 114} (2015), no.~11 111601,
  [\href{http://xxx.lanl.gov/abs/1312.0999}{{\tt arXiv:1312.0999}}].

\bibitem{Polchinski:1991ax}
J.~Polchinski and A.~Strominger, {\it {Effective string theory}},  {\em
  Phys.Rev.Lett.} {\bf 67} (1991) 1681--1684.

\bibitem{Aharony:2009gg}
O.~Aharony and E.~Karzbrun, {\it {On the effective action of confining
  strings}},  {\em JHEP} {\bf 0906} (2009) 012,
  [\href{http://xxx.lanl.gov/abs/0903.1927}{{\tt arXiv:0903.1927}}].

\bibitem{ASY}
O.~Aharony, J.~Sonnenschein, and S.~Yankielowicz, {\it {On the quantization of
  rotating open strings with massive endpoints [In preparation]}}, .

\bibitem{Kinar:1998vq}
Y.~Kinar, E.~Schreiber, and J.~Sonnenschein, {\it {Q anti-Q potential from
  strings in curved space-time: Classical results}},  {\em Nucl.Phys.} {\bf
  B566} (2000) 103--125, [\href{http://xxx.lanl.gov/abs/hep-th/9811192}{{\tt
  hep-th/9811192}}].

\bibitem{Kruczenski:2004me}
M.~Kruczenski, L.~A. Pando~Zayas, J.~Sonnenschein, and D.~Vaman, {\it {Regge
  trajectories for mesons in the holographic dual of large-N(c) QCD}},  {\em
  JHEP} {\bf 0506} (2005) 046,
  [\href{http://xxx.lanl.gov/abs/hep-th/0410035}{{\tt hep-th/0410035}}].

\bibitem{Bigazzi:2004ze}
F.~Bigazzi, A.~Cotrone, L.~Martucci, and L.~Pando~Zayas, {\it {Wilson loop,
  Regge trajectory and hadron masses in a Yang-Mills theory from semiclassical
  strings}},  {\em Phys.Rev.} {\bf D71} (2005) 066002,
  [\href{http://xxx.lanl.gov/abs/hep-th/0409205}{{\tt hep-th/0409205}}].

\bibitem{Bali:2000un}
G.~S. Bali, {\it {Casimir scaling of SU(3) static potentials}},  {\em
  Phys.Rev.} {\bf D62} (2000) 114503,
  [\href{http://xxx.lanl.gov/abs/hep-lat/0006022}{{\tt hep-lat/0006022}}].

\bibitem{Armoni:2006ri}
A.~Armoni and B.~Lucini, {\it {Universality of k-string tensions from
  holography and the lattice}},  {\em JHEP} {\bf 06} (2006) 036,
  [\href{http://xxx.lanl.gov/abs/hep-th/0604055}{{\tt hep-th/0604055}}].

\bibitem{Donnachie:1984xq}
A.~Donnachie and P.~Landshoff, {\it {Elastic Scattering and Diffraction
  Dissociation}},  {\em Nucl.Phys.} {\bf B244} (1984) 322.

\bibitem{Isgur:1984bm}
N.~Isgur and J.~E. Paton, {\it {A Flux Tube Model for Hadrons in QCD}},  {\em
  Phys.Rev.} {\bf D31} (1985) 2910.

\bibitem{Meyer:2004jc}
H.~B. Meyer and M.~J. Teper, {\it {Glueball Regge trajectories and the pomeron:
  A Lattice study}},  {\em Phys.Lett.} {\bf B605} (2005) 344--354,
  [\href{http://xxx.lanl.gov/abs/hep-ph/0409183}{{\tt hep-ph/0409183}}].

\bibitem{Meyer:2004gx}
H.~B. Meyer, {\it {Glueball regge trajectories}},
  \href{http://xxx.lanl.gov/abs/hep-lat/0508002}{{\tt hep-lat/0508002}}.

\bibitem{Burden:1982zb}
C.~Burden and L.~Tassie, {\it {Rotating Strings, Glueballs and Exotic Mesons}},
   {\em Austral.J.Phys.} {\bf 35} (1982) 223--233.

\bibitem{'tHooft:2004he}
G.~'t~Hooft, {\it {Minimal strings for baryons}},
  \href{http://xxx.lanl.gov/abs/hep-th/0408148}{{\tt hep-th/0408148}}.

\bibitem{Cotrone:2005fr}
A.~L. Cotrone, L.~Martucci, and W.~Troost, {\it {String splitting and strong
  coupling meson decay}},  {\em Phys. Rev. Lett.} {\bf 96} (2006) 141601,
  [\href{http://xxx.lanl.gov/abs/hep-th/0511045}{{\tt hep-th/0511045}}].

\bibitem{Bigazzi:2006jt}
F.~Bigazzi and A.~L. Cotrone, {\it {New predictions on meson decays from string
  splitting}},  {\em JHEP} {\bf 11} (2006) 066,
  [\href{http://xxx.lanl.gov/abs/hep-th/0606059}{{\tt hep-th/0606059}}].

\bibitem{Hashimoto:2007ze}
K.~Hashimoto, C.-I. Tan, and S.~Terashima, {\it {Glueball decay in holographic
  QCD}},  {\em Phys. Rev.} {\bf D77} (2008) 086001,
  [\href{http://xxx.lanl.gov/abs/0709.2208}{{\tt arXiv:0709.2208}}].

\bibitem{Brunner:2015oqa}
F.~Br{\"u}nner, D.~Parganlija, and A.~Rebhan, {\it {Glueball Decay Rates in the
  Witten-Sakai-Sugimoto Model}},  {\em Phys. Rev.} {\bf D91} (2015), no.~10
  106002, [\href{http://xxx.lanl.gov/abs/1501.0790}{{\tt arXiv:1501.0790}}].

\bibitem{Brunner:2015yha}
F.~Br{\"u}nner and A.~Rebhan, {\it {Nonchiral enhancement of scalar glueball
  decay in the Witten-Sakai-Sugimoto model}},
  \href{http://xxx.lanl.gov/abs/1504.0581}{{\tt arXiv:1504.0581}}.

\bibitem{Crede:2008vw}
V.~Crede and C.~Meyer, {\it {The Experimental Status of Glueballs}},  {\em
  Prog.Part.Nucl.Phys.} {\bf 63} (2009) 74--116,
  [\href{http://xxx.lanl.gov/abs/0812.0600}{{\tt arXiv:0812.0600}}].

\bibitem{PDG:2014}
{\bf Particle Data Group} Collaboration, K.~Olive et~al., {\it {Review of
  Particle Physics}},  {\em Chin.Phys.} {\bf C38} (2014) 090001.

\bibitem{Klempt:2007cp}
E.~Klempt and A.~Zaitsev, {\it {Glueballs, Hybrids, Multiquarks. Experimental
  facts versus QCD inspired concepts}},  {\em Phys.Rept.} {\bf 454} (2007)
  1--202, [\href{http://xxx.lanl.gov/abs/0708.4016}{{\tt arXiv:0708.4016}}].

\bibitem{Mathieu:2008me}
V.~Mathieu, N.~Kochelev, and V.~Vento, {\it {The Physics of Glueballs}},  {\em
  Int.J.Mod.Phys.} {\bf E18} (2009) 1--49,
  [\href{http://xxx.lanl.gov/abs/0810.4453}{{\tt arXiv:0810.4453}}].

\bibitem{Ochs:2013gi}
W.~Ochs, {\it {The Status of Glueballs}},  {\em J.Phys.} {\bf G40} (2013)
  043001, [\href{http://xxx.lanl.gov/abs/1301.5183}{{\tt arXiv:1301.5183}}].

\bibitem{Nebreda:2011cp}
J.~Nebreda, J.~Pelaez, and G.~Rios, {\it {Enhanced non-quark-antiquark and
  non-glueball Nc behavior of light scalar mesons}},  {\em Phys.Rev.} {\bf D84}
  (2011) 074003, [\href{http://xxx.lanl.gov/abs/1107.4200}{{\tt
  arXiv:1107.4200}}].

\bibitem{Anisovich:2000kxa}
A.~Anisovich, V.~Anisovich, and A.~Sarantsev, {\it {Systematics of q anti-q
  states in the (n, M**2) and (J, M**2) planes}},  {\em Phys.Rev.} {\bf D62}
  (2000) 051502, [\href{http://xxx.lanl.gov/abs/hep-ph/0003113}{{\tt
  hep-ph/0003113}}].

\bibitem{Anisovich:2002us}
V.~Anisovich, {\it {Systematics of quark anti-quark states and scalar exotic
  mesons}},  {\em Phys.Usp.} {\bf 47} (2004) 45--67,
  [\href{http://xxx.lanl.gov/abs/hep-ph/0208123}{{\tt hep-ph/0208123}}].

\bibitem{Masjuan:2012gc}
P.~Masjuan, E.~Ruiz~Arriola, and W.~Broniowski, {\it {Systematics of radial and
  angular-momentum Regge trajectories of light non-strange
  $q\overline{q}$-states}},  {\em Phys.Rev.} {\bf D85} (2012) 094006,
  [\href{http://xxx.lanl.gov/abs/1203.4782}{{\tt arXiv:1203.4782}}].

\bibitem{Bugg:2012yt}
D.~Bugg, {\it {Comment on “Systematics of radial and angular-momentum Regge
  trajectories of light nonstrange $q\overline{q}$-states”}},  {\em
  Phys.Rev.} {\bf D87} (2013), no.~11 118501,
  [\href{http://xxx.lanl.gov/abs/1209.3481}{{\tt arXiv:1209.3481}}].

\bibitem{Ablikim:2012ft}
{\bf BESIII} Collaboration, M.~Ablikim et~al., {\it {Study of the
  near-threshold ωϕ mass enhancement in doubly OZI-suppressed J/ψ→γωϕ
  decays}},  {\em Phys.Rev.} {\bf D87} (2013), no.~3 032008,
  [\href{http://xxx.lanl.gov/abs/1211.5668}{{\tt arXiv:1211.5668}}].

\bibitem{Bai:1996wm}
{\bf BES Collaboration} Collaboration, J.~Bai et~al., {\it {Studies of xi
  (2230) in J / psi radiative decays}},  {\em Phys.Rev.Lett.} {\bf 76} (1996)
  3502--3505.

\bibitem{Benslama:2002pa}
{\bf CLEO} Collaboration, K.~Benslama et~al., {\it {Anti-search for the
  glueball candidate f(J)(2220) in two -photon interactions}},  {\em Phys.Rev.}
  {\bf D66} (2002) 077101, [\href{http://xxx.lanl.gov/abs/hep-ex/0204019}{{\tt
  hep-ex/0204019}}].

\bibitem{Vladimirsky:2001ek}
V.~Vladimirsky, V.~Grigorev, O.~Erofeeva, Y.~Katinov, V.~Lisin, et~al., {\it
  {Resonance maximum in the system of two K(S) mesons at 1450-MeV}},  {\em
  Phys.Atom.Nucl.} {\bf 64} (2001) 1895--1897.

\bibitem{Vijande:2004he}
J.~Vijande, F.~Fernandez, and A.~Valcarce, {\it {Constituent quark model study
  of the meson spectra}},  {\em J.Phys.} {\bf G31} (2005) 481,
  [\href{http://xxx.lanl.gov/abs/hep-ph/0411299}{{\tt hep-ph/0411299}}].

\bibitem{Athenodorou:2010cs}
A.~Athenodorou, B.~Bringoltz, and M.~Teper, {\it {Closed flux tubes and their
  string description in D=3+1 SU(N) gauge theories}},  {\em JHEP} {\bf 02}
  (2011) 030, [\href{http://xxx.lanl.gov/abs/1007.4720}{{\tt
  arXiv:1007.4720}}].

\bibitem{Bochicchio:2013aha}
M.~Bochicchio, {\it {Yang-Mills mass gap, Floer homology, glueball spectrum,
  and conformal window in large-N QCD}},
  \href{http://xxx.lanl.gov/abs/1312.1350}{{\tt arXiv:1312.1350}}.

\bibitem{Bochicchio:2013sra}
M.~Bochicchio, {\it {Glueball and meson spectrum in large-N massless QCD}},
  \href{http://xxx.lanl.gov/abs/1308.2925}{{\tt arXiv:1308.2925}}.

\bibitem{Caselle:2015tza}
M.~Caselle, A.~Nada, and M.~Panero, {\it {Hagedorn spectrum and thermodynamics
  of SU(2) and SU(3) Yang-Mills theories}},
  \href{http://xxx.lanl.gov/abs/1505.0110}{{\tt arXiv:1505.0110}}.

\bibitem{Gregory:2012hu}
E.~Gregory, A.~Irving, B.~Lucini, C.~McNeile, A.~Rago, et~al., {\it {Towards
  the glueball spectrum from unquenched lattice QCD}},  {\em JHEP} {\bf 1210}
  (2012) 170, [\href{http://xxx.lanl.gov/abs/1208.1858}{{\tt
  arXiv:1208.1858}}].

\bibitem{Albanese:1987ds}
{\bf APE} Collaboration, M.~Albanese et~al., {\it {Glueball Masses and String
  Tension in Lattice QCD}},  {\em Phys.Lett.} {\bf B192} (1987) 163--169.

\bibitem{Morningstar:1999rf}
C.~J. Morningstar and M.~J. Peardon, {\it {The Glueball spectrum from an
  anisotropic lattice study}},  {\em Phys.Rev.} {\bf D60} (1999) 034509,
  [\href{http://xxx.lanl.gov/abs/hep-lat/9901004}{{\tt hep-lat/9901004}}].

\bibitem{Chen:2005mg}
Y.~Chen, A.~Alexandru, S.~Dong, T.~Draper, I.~Horvath, et~al., {\it {Glueball
  spectrum and matrix elements on anisotropic lattices}},  {\em Phys.Rev.} {\bf
  D73} (2006) 014516, [\href{http://xxx.lanl.gov/abs/hep-lat/0510074}{{\tt
  hep-lat/0510074}}].

\bibitem{Bali:1993fb}
{\bf UKQCD Collaboration} Collaboration, G.~Bali et~al., {\it {A Comprehensive
  lattice study of SU(3) glueballs}},  {\em Phys.Lett.} {\bf B309} (1993)
  378--384, [\href{http://xxx.lanl.gov/abs/hep-lat/9304012}{{\tt
  hep-lat/9304012}}].

\bibitem{Lucini:2014paa}
B.~Lucini, {\it {Glueballs from the Lattice}},  {\em PoS} {\bf QCD-TNT-III}
  (2013) 023, [\href{http://xxx.lanl.gov/abs/1401.1494}{{\tt
  arXiv:1401.1494}}].

\bibitem{Lucini:2004my}
B.~Lucini, M.~Teper, and U.~Wenger, {\it {Glueballs and k-strings in SU(N)
  gauge theories: Calculations with improved operators}},  {\em JHEP} {\bf
  0406} (2004) 012, [\href{http://xxx.lanl.gov/abs/hep-lat/0404008}{{\tt
  hep-lat/0404008}}].

\bibitem{Ganor:1994rm}
O.~Ganor, J.~Sonnenschein, and S.~Yankielowicz, {\it {Folds in 2-D string
  theories}},  {\em Nucl.Phys.} {\bf B427} (1994) 203--244,
  [\href{http://xxx.lanl.gov/abs/hep-th/9404149}{{\tt hep-th/9404149}}].

\end{thebibliography}\endgroup
\end{document}